\documentclass[12pt]{article}
\usepackage{a4wide}
\usepackage{psfrag}
\usepackage{graphics}
\usepackage{amsfonts}
\usepackage{amstext}
\usepackage{amsmath}
\usepackage{amssymb}
\usepackage{amsthm}
\usepackage[mathscr]{eucal}
\makeatletter
 
 \@addtoreset{equation}{section}
\makeatother

\def\Ai{{\rm Ai}}

\def\e{\epsilon}
\def\g{\gamma}
\def\G{\mathcal{G}}

\def\K{\mathcal{K}}

\def\sh{\sharp}

\def\t{\tau}

\begin{document}
\title{\bf Dynamics of a tagged particle in the asymmetric exclusion 
process \\with the step initial condition}

\author{
\vspace{5mm}
T. Imamura
{\footnote {\tt e-mail: timamura@iis.u-tokyo.ac.jp}}
~~and T. Sasamoto 
{\footnote {\tt e-mail: sasamoto@math.s.chiba-u.ac.jp}}
\\
{\it $^*$Institute of Industrial Science, University of Tokyo,}\\
\vspace{5mm}
{\it Komaba 4-6-1, Meguro-ku, Tokyo 153-8505, Japan}\\
{\it $^{\dag}$Department of Mathematics and Informatics,}\\
{\it Chiba University,}\\
{\it Yayoi-cho 1-33, Inage-ku, Chiba 263-8522, Japan}\\
}
\date{}
\maketitle

\begin{abstract}
The one-dimensional totally asymmetric simple exclusion process (TASEP) 
is considered. We study the time evolution property of a tagged 
particle in the TASEP with the step initial condition. Calculated is 
the multi-time joint distribution function of its position. Using the 
relation of the dynamics of the TASEP to the Schur process, we show that 
the function is represented as the Fredholm determinant. We also study 
the scaling limit. The universality of the largest eigenvalue in the 
random matrix theory is realized in the limit. When the hopping rates 
of all particles are the same, it is found that the joint distribution 
function converges to that of the Airy process after the time at which
the particle begins to move. On the other hand, when there are several
particles with small hopping rate in front of a tagged particle, the 
limiting process changes at a certain time from the Airy process to the 
process of the largest eigenvalue in the Hermitian multi-matrix model 
with external sources.

\vspace{3mm}\noindent
[Keywords: asymmetric simple exclusion process; KPZ universality 
class; random matrices; Tracy-Widom distribution; Airy process]
\end{abstract}

\newpage
\section{Introduction}
Dynamics of a nonequilibrium system is one of the most stimulating 
topics in statistical mechanics. The dynamical property is determined 
by the interplay among various elements such as interaction, 
initial and boundary conditions, and so on. But little is known 
about how it depends on these elements and realizes universality 
in a situation where the elements are intertwined. The aim of 
this article is to approach such questions by analyzing the dynamics 
of the one-dimensional asymmetric simple exclusion process (ASEP).   

The ASEP is one of the most typical models of interacting particle 
processes where particles diffuse to a preferred direction with 
hard core repulsive interaction~\cite{Li1985, Li1999, Sp1991, 
ScZi1994, Sc2001}. In spite of the simplicity of the model, it has 
been known that the ASEP shows various interesting phenomena caused by 
the collaboration between the diffusion and exclusion effect. 
Furthermore, in the one-dimensional case, it has an integrable 
mathematical structure which allows us to analyze some physical 
quantities exactly. For example, in the open boundary condition a 
steady state can be constructed using the matrix product method 
and $q$-orthogonal polynomials, and the boundary-induced phase 
transition can be discussed~\cite{DEHP1993, Sa1999, USW2004}.

On the other hand, the dynamical (non-stationary) properties of the 
one-dimensional ASEP have been also analyzed recently. Among various 
studies on the topic, we focus on the problem of the diffusion of 
a particular particle in the one-dimensional ASEP. (We call the
particle a {\it{tagged}} particle.) This is a fundamental and 
elementary problem for the understanding of the dynamics of the ASEP. We 
are interested in how initial conditions and the exclusion effect 
through hard-core repulsive interaction affect the diffusion property 
and how it is different from the (normal) diffusion of the Brownian 
particle.

The scaling exponent of the diffusion provides insights into these 
problems. It was found that it depends on the initial conditions. 
In a steady state with a given density, the position fluctuation of 
a tagged particle grows as $t^{1/2}$~\cite{Ki1986}. The exponent 1/2 
is the same as that of the Brownian particle.  On the other hand, 
for the fixed initial conditions where initially we create particles 
with a given density and fix the initial configuration, the exponent 
changes from 1/2 to 1/3~\cite{vB1991,MB1991}.  Such a diffusion is 
called an anomalous diffusion. This exponent 1/3 can be understood 
from the fact that the time evolution of the one-dimensional ASEP can 
be described by the one-dimensional Kardar-Parisi-Zhang universality 
class~\cite{KPZ1986}. In the study of the current in the 
one-dimensional ASEP on a ring, the exponent 1/3 was calculated by the 
Bethe ansatz technique~\cite{GS1992,Ki1995}.

Note that the two cases above have a common property that the 
density is invariant under both time and space translation 
and a tagged particle moves with a fixed velocity on average.
On the other hand, let us now consider another initial condition 
where all sites on the left of a certain bond are occupied
while all right sites are empty. We call this initial condition
a {\it {step}} initial condition. This is the typical initial
condition where the density and the average velocity of a tagged 
particle change with time.  Under the initial condition, we focus 
on the $M$th particle from the right as a tagged particle 
and consider the time evolution of the particle when $M$ is large
(but finite). In order to study this problem, we investigate
the limiting behavior of the diffusion of the tagged particle 
as both $M$ and time go to infinity. In the totally asymmetric
simple exclusion process (TASEP), where particles move only to  
the right, the average position of a tagged particle has been
given in the scaling limit~\cite{Se1997}. On the other hand, 
we are interested in the limiting process itself which the tagged 
particle obeys, as well as the average position and the scaling 
exponent. 

Such an attempt as getting more sophisticated information than the 
scaling exponent has been recently discussed in the study on 
the current fluctuations in the TASEP. Not only the exponent but also the 
scaling function were obtained and  it was revealed that the 
function is equivalent to the largest eigenvalue distribution in the 
random matrix theory with various universality classes. This development 
is based on the results on the longest increasing subsequence in a random 
permutation~\cite{BDJ1999, Jo2000,BR2001a,BR2001b,BR2001c}.

Remarkable is that the scaling function detects a difference 
in initial conditions even if the scaling exponent does not. 
For instance, in the case of the step initial condition, 
the scaling function is equivalent to the GUE Tracy-Widom 
distribution~\cite{TW1994}, which is the largest eigenvalue 
distribution of the Gaussian unitary ensemble (GUE)~\cite{Jo2000, 
NS2004, RS2005}. In other initial conditions, on the other hand, these 
are described as the largest eigenvalue distributions with other 
universality classes in RMT such as Gaussian orthogonal ensemble 
(GOE)~\cite{TW1996}, two independent GOE's which is denoted by 
GOE$^2$~\cite{BR2001b,BR2001c}, and so on~\cite{NS2004,PS2002a}. In 
addition, the equal-time multipoint distribution of the current 
fluctuations is obtained in the step and periodic initial 
conditions~\cite{Sa2005} by use of the Green function of 
the TASEP~\cite{S1997}.

In this article, we calculate the multi-time distribution function 
of position fluctuations of a tagged particle in the TASEP with the
step initial condition based on the techniques of the directed 
polymer problem in a 01 matrix~\cite{Se1998,J2001,GTW2001, GTW2002a} 
and stochastic growth of Young diagram characterized by the Schur 
process~\cite{OC2003a, OC2003b,BO2006a,BO2006b, OkRe2003, BoRa2006}. 
We express the function as the Fredholm determinant and discuss the
scaling limit by the saddle point analysis. 
We remark that what will be considered in this paper is 
the pointwise convergence of the kernel. The convergence of 
the Fredholm determinant itself is certainly expected to hold 
from previous works on related problems \cite{Jo2003,BFP2006p}, 
but more elaborate asymptotic analysis is necessary for its proof.
In the study on the ASEP, such correlations between different times 
have not been studied yet although the Green function~\cite{RS2005,
S1997,PP2006} and the equal-time joint distributions of particle 
positions~\cite{BFP2006p,BFPS2006p} have been studied recently. In 
the one-time case, on the other hand, the position fluctuation is 
essentially the same as the current fluctuation which has already been 
studied in~\cite{Jo2000,PS2002a}. But it has not been stated as the 
position fluctuation.  

Furthermore, we also consider the dependence of the distribution 
function on defect particles located in front of the tagged particle.  
Here defect particle means a particle with different hopping 
rate from other normal particles and we assume the number of them 
is finite. Such a situation as the TASEP with particle-dependent hopping 
rates has been recently discussed in~\cite{BBAP2005, Ba2006, RS2006}.
We clarify that when their hopping rate is smaller than that of normal 
particles, the limiting behavior of the multi-time distribution changes 
due to their presence although their number is finite. 

The paper is organized as follows. In the next section, we describe
the definition of the model. The main result of this article is
given in Section~\ref{main}. In Section~\ref{ftts}, we express the 
multi-time distribution of position fluctuations in the TASEP as Fredholm 
determinant. For this purpose, we first show that the directed polymer 
problem on a 01 matrix is related to the time evolution of a tagged 
particle in the TASEP with the step initial condition. Next, by 
applying  the dual Robinson-Schensted-Knuth algorithm to the 01 
matrix, we map the tagged particle problem in the TASEP to the stochastic 
growth of Young diagram described by the Schur process.  In 
Section~\ref{asymp}, we discuss the two types of scaling limit for the 
multi-time distribution function, the case where both time $t$ and the 
label of the tagged particle $M$ tend to infinity, and the one where $t$ 
goes to infinity with $M$ fixed. Some discussions and concluding remark 
are given in Section~\ref{discussion} and Section~\ref{conclusion} 
respectively. 

\section{Model}\label{model}
In this section, we define the model and the quantity which we study in 
this article. Let us consider the one-dimensional infinite lattice and 
particles as illustrated in Fig.~1. Each site can be occupied by at most 
one particle. Suppose all particles are labeled an integer $i$ from the 
right. A particle moves stochastically obeying the following rules. 
During each time step between $t\in \{0,1,\cdots\}$ and $t+1$, the 
particle labeled $i$ hops to the right neighboring site with probability 
$1-q_i$ ($0\leq q_i < 1$) and stays at the same site with probability 
$q_i$ if the right neighboring site is empty. On the other hand, if 
the site is occupied by the particle labeled $i-1$, the $i$th particle 
stays at the same site with probability 1. This incorporates the 
exclusion effect which describes the hard core repulsive interaction. 
When hopping rates of all particles are the same, i.e., $q_i = q 
(\forall i)$, the system is the usual totally asymmetric simple 
exclusion process (TASEP) with parallel update. We allow the 
particle-dependent hopping rates and study the effects of defect 
particles. 

In this article, we consider the step initial condition in which
all sites on the left of a certain bond between two sites are occupied 
by particles and all sites on the right of the bond are empty. Fig.~1(b) 
illustrates the initial condition.

Under these settings, we label the particles at $1,2,\cdots$ from the 
right (see Fig.~1(b)) and consider the dynamics of the particle labeled 
$M$. (This particle which we focus on is called the tagged particle.)
For this purpose, we set the position coordinate such that the tagged 
particle is at the origin at time 0 as depicted in Fig.~1(b). Let us 
define $L(t,M)$ as the position of the $M$th particle at time $t$. In 
other words, $L(t,M)$ represents the distance travelled by the $M$th 
particle from time $0$ through $t$.

The quantity which we will investigate is the multi-time joint 
distribution function for $L(t,M)$ defined as
\begin{equation}
\text{Prob}\left(L(t_1,M)\ge\ell_1,L(t_2,M)\ge\ell_2,\cdots,
L(t_m,M)\ge\ell_m\right).
\label{1}
\end{equation}
In particular, we are interested in the asymptotic behavior of~\eqref{1}
as $t$ goes to infinity. In this article, two types of scaling limit 
will be discussed. The first one is the limit where both time $t$ and 
the label of the tagged particle $M$ go to infinity with the ratio 
$t/M$ fixed. In the second one, on the other hand, we take the 
$t\to\infty$ limit with $M$ fixed.  

In order to see how the whole particles from the first to $M$th travel, 
we performed the numerical simulation of which the result is shown 
in Fig.~2. We considered the time evolution of the particles from the 
first to 100th and assumed the hopping rate of all particles is 0.9 
except the first, 25th, 50th and 75th particles whose hopping rate is 
0.8. Thus these four particles represent the defect particles. The 
typical example of the time evolution of the 100 particles from time 
0 through 3000 is illustrated in Fig.~2(a).  In this figure, the 
vertical axis represents the position coordinate introduced 
in Fig.~1(b), whereas the horizontal axis is time axis, and each 
particle is expressed as $+$. From the macroscopic point of view such 
as Fig.~2(a), we only find that they travel in a group. However, 
if we zoom the configuration of the particles in a certain time zone, 
we can see the microscopic pattern of the configuration. Here we show 
two characteristic configurations which are depicted in Fig.~2(b) and 
2(c). Fig.~2(b) (resp.(c)) represents the configuration of the 
particles  when time is around 200 (resp. 3000). When the time zone 
which we focus on is early enough as in Fig.~2(b), the particles move 
forming only one group and the four defect particles are included in 
the group. On the other hand, when time has passed sufficiently as 
in Fig.~2(c), they form the four groups automatically and the top of 
each group is the defect particle. The similar platoon structure was 
also discussed in the study of the TASEP with disordered hopping 
rates~\cite{E1996,E1997,SeKr1999}. In fact, as will be explained in 
the next section, the position fluctuation of a tagged particle also 
changes with time corresponding to the change of the configuration.

\section{Main results}\label{main}
\subsection{Multi-time distributions}\label{multitimed}
Our whole discussions in this article are based on a fact that the 
tagged particle problem for the step initial condition has a nice 
combinatorial structure which admits us to obtain a closed expression 
for the multi-time joint distribution in the form of the Fredholm 
determinant. The results are summarized in the following theorem.
The proof will be presented in Section~\ref{ftts}.

Note that due to the rule of the TASEP, the particle labeled $M$ cannot
move when $t<M$. Thus, throughout the article, we consider the multi-time 
function~\eqref{1} under the condition
\begin{equation}
t_i\ge M.
\label{1cond}
\end{equation}
\vspace{3mm}

\noindent
{\bf Theorem 1}
{\it

When \eqref{1cond} holds,
\begin{equation}
{\rm Prob}\left(L(t_1,M)\ge\ell_1,L(t_2,M)\ge\ell_2,\cdots,
L(t_m,M)\ge\ell_m\right)
=\det\left(1+Kg\right).
\end{equation}
Here $\det(1+Kg)$ is the Fredholm determinant defined as
\begin{align}
&~\det\left(1+Kg\right)\notag\\
&=\sum_{k=0}^{\infty}\frac{1}{k!}\sum_{n_1=1}^{m}\sum_{x_1=-\infty}^{\infty}
\cdots\sum_{n_k=1}^{m}\sum_{x_k=-\infty}^{\infty}
g(t_{n_1};x_1)\cdots
g(t_{n_k};x_k)
\det\left(K(t_{n_l},x_l;t_{n_{l'}},x_{l'})\right)_{l,l'=1}^k,
\label{fFreddef}
\end{align}
where
\begin{align}
g(t_n;x)
&=
-\chi_{(t_n-M+1-\ell_n,\infty)}(x),\quad (n=1,2,\ldots,m),
\notag\\
\chi_{(a,b)}(x)
&=
\begin{cases}
1,& \text{if~~} a< x< b,\\
0, & \text{otherwise},
\end{cases}
\label{chi}
\end{align}
and the kernel $K(t_1,x_1;t_2,x_2)$ is given by
\begin{align}
&~K(t_1,x_1;t_2,x_2)=\tilde{K}(t_1,x_1;t_2,x_2)
-\phi_{t_1,t_2}(x_1,x_2),
\label{kernel}
\\
&~\tilde{K}(t_1,x_1;t_2,x_2)\notag\\
&=\frac{1}{(2\pi i)^2}\int_{C_{R_1}}\frac{dz_1}{z_1}\int_{C_{R_2}}
\frac{dz_2}{z_2}\frac{z_1}{z_1-z_2}
\frac{(1+1/z_2)^{t_2-M+1}}{(1+1/z_1)^{t_1-M+1}}\prod_{i=1}^M
\frac{1-q_i-q_iz_2}{1-q_i-q_iz_1}\frac{z_2^{x_2}}{z_1^{x_1}},
\label{Ktxtx}
\\
&\phi_{t_1,t_2}(x_1,x_2)=
\begin{cases}
\frac{1}{2\pi i}\int_{C_1} \frac{dz}{z}
\left(1+\frac{1}{z}\right)^{t_2-t_1}z^{x_2-x_1},&
t_1<t_2,\\
0,& t_1\ge t_2. 
\end{cases}
\label{thm1f}
\end{align}
Here $C_{R}$ in~\eqref{Ktxtx} and~\eqref{thm1f} denotes a contour 
enclosing the origin anticlockwise with radius $R$ and $R_{i}~(i=1,2)$  
in~\eqref{Ktxtx} satisfy the conditions $R_2<R_1$ and 
$1<R_1<(1-q_i)/q_i$.  
}
\vspace{3mm}

\subsection{Scaling limit 1 ($M\to\infty$)}\label{sclim1}
Using the results in Theorem 1, one can study the asymptotics of the 
joint distribution. In this subsection, we consider the scaling limit 
such that both $t$ and $M$ go to infinity with their ratio
\begin{equation}
u = t/M
\label{1sl}
\end{equation}
fixed. $u$ represents the scaled time. Notice that we also take the 
limit of the label of the tagged particle $M$. This may sound strange 
if the particle we are focusing on is varied as $M$ goes to infinity. 
Rather we think that the results below give the asymptotic behaviors
of the path statistics of the tagged particle for large but finite $M$
which is fixed. Alternatively one might be able to interpret them 
as giving the asymptotic behaviors of the path statistics of the tagged 
particle near a large but finite $t$.

We also discuss the effect of defect particles. In particular, we 
consider the situation where there are finite $n$ defect particles 
with stay rates $\bar{q}_i(i=1,\cdots,n)$ in front of the tagged 
particle. For this purpose, we set the stay rate $q_i$ of $i$th 
particle as follows. For some set $\{a_i\}_{i=1,\cdots,n}\subset
\{1,2,\cdots M\}$ with $n$ fixed, we assume 
\begin{equation}
q_{i}=
\begin{cases}
\bar{q}_j,& {\text{if~~}} i=a_j,\\
q,  & {\text{otherwise}}.
\end{cases}
\label{1dp} 
\end{equation}
Note that the particle with the rate $\bar{q}_j$ represents the defect 
particle whose label is $a_j$ whereas the remaining particles with the
rate $q$ are the normal particles. 

\subsubsection{Average position}
The average position of the tagged particle divided by $M$ has a 
deterministic limit as $M\to\infty$. Let us call it $A(u)$,
\begin{equation}
A(u)=\lim_{M\rightarrow\infty}\frac{L(t=uM,M)}{M}.
\label{avpar}
\end{equation}
We also define $\bar{q}=\max\{\bar{q}_i\}$. (Note that $\bar{q}$ is the stay 
rate of the slowest defect particle.) If $\bar{q}>q$, which is 
the case where the slowest defect particle is slower than the normal
particles, we obtain 
\begin{equation}
A(u)
=
\begin{cases}
0,& \text{if~~}u\le \frac{1}{1-q},\\
A_2(u),&
\text{if~~}\frac{1}{1-q}\le u\le u_c,\\
A_G(u),&\text{if~~} u_c\le u,
\end{cases}
\label{average}
\end{equation}
where $u_c=(\bar{q}^2-2q\bar{q}+q)/{(\bar{q}-q)^2}$ and
\begin{align}
A_2(u)&=(1-q)u-(1-2q)-2\sqrt{q(1-q)(u-1)},
\label{A2}
\\
A_G(u)&=(1-\bar{q})u-\frac{(1-\bar{q})\bar{q}}
{\bar{q}-q}.
\label{AG}
\end{align}
$A_2(u)$ can be understood from Theorem 1.1. in~\cite{Jo2000}.
If $\bar{q}\le q$, the average position is represented as only the 
first and second cases in~\eqref{average}. ($u_c$ and $A_G(u)$ do not 
appear in the case.) Note that $A_2(u)$ does not depend on the stay 
rate $\bar{q}$ of the slowest defect particle whereas $A_G(u)$ does. 
$A(u)$ is illustrated in Fig.~3.

Next we consider the position fluctuations of the tagged particle 
around the average position. There are four typical regions shown in 
Fig.~3. In each region, we can obtain the result on the position 
fluctuations by focusing on a point within the region and taking a 
proper scaling around the point. Note that in Fig.~3, the regions 2 and 
4 spread in the form of a line whereas the regions 1 and 3 are 
point-like.  Thus, in the region 1(resp. 3), we always consider the 
fluctuation property around the point $1/(1-q)$ (resp. $u_c=(\bar{q}^2
-2q\bar{q}+q)/(\bar{q}-q)^2$).

\subsubsection{Region 1 ($u=1/(1-q)$)} 
This is the region where the tagged particle just starts to move. This 
corresponds to $u=1/(1-q)$. We set
\begin{align}
&t_i=\frac{M}{1-q}+D_1 M^{\frac12}\t_i,
\label{r1ti}
\end{align}
where $D_1=\sqrt{q}/(1-q)$. Note that the time $t_i$~\eqref{r1ti} is 
approximately equal to the time on which some leftmost holes arrive at 
the site occupied by the tagged particle. One easily finds that the 
arrival time can be scaled as in~\eqref{r1ti} which is the same scaling 
as the central limit theorem. 

Applying the scaling to Theorem 1,  we have the theorem as follows. 
The proof will be given in Section~\ref{pot21}.

\vspace{3mm}
\noindent
{\bf Theorem 2-1}
{\it
\begin{equation}
\lim_{M\rightarrow\infty}{\rm Prob}
\left(L(t_1,M)\ge\ell_1,L(t_2,M)\ge\ell_2,\cdots,
L(t_m,M)\ge\ell_m\right)=\det\left(1+\K_{1}g_1\right).
\end{equation}
Here the Fredholm determinant in the right hand side is defined in
\eqref{fFreddef} where $g_1(\t_i,x)=-\chi_{(-\infty,\ell_i)}(x)$. 
$\chi_{(a,b)}(x)$ is defined in~\eqref{chi}. The kernel is given by
\begin{align}
\K_{1}(\t_1,x_1;\t_2,x_2)
=
\begin{cases}
\sum_{m=0}^{\infty}\psi_1(x_1-m,\t_1)\psi_{2}(x_2-m,\t_2),&
\t_1\ge \t_2,\\
-\sum_{m=0}^{\infty}\psi_1(x_1+m+1,\t_1)\psi_{2}(x_2+m+1,\t_2),
&\t_1<\t_2.
\end{cases}
\label{th1-1}
\end{align}
Here $\psi_1(x_1,\t_1)$ can be represented by use of
the parabolic cylinder function $D_n(x)$ 
in~\cite{AAR1999} as
\begin{equation}
\psi_1(x_1,\t_1)=\frac{1}{2\pi i}\int_{-i\infty+\e}^{i\infty+\e}
dz{e^{\frac{z^2}{2}-\t_1z}}{z^{x_1-1}}
=\frac{e^{-\frac{x_1^2}{4}}}{\sqrt{2\pi}}D_{x_1-1}(\t),
\label{r1psi1}
\end{equation}
where $\e$ is taken to be positive, and 
\begin{align}
\psi_2(x_2,\t_2)=\frac{1}{2\pi i}\int_{C_1} \frac{dw}{w^{x_2+1}} 
e^{-\frac{w^2}{2}+\t_2w}
=
\begin{cases}
\frac{2^{\frac{-x_2}{2}}}{x_2!}H_{x_2}
\left(\frac{\t}{\sqrt{2}}\right), 
& x_2 \ge 0,\\
0,& x_2<0,
\end{cases}
\label{th1-3}
\end{align}
where $C_1$ in~\eqref{th1-3} represents the contour enclosing 
the origin anticlockwise with radius 1, and $H_{n}(x)$ is the 
Hermite polynomial with degree $n$~\cite{AAR1999}. 
}
\vspace{3mm}

The special case where $\t_1=\cdots=\t_m=0$ has appeared in the 
distribution of the height fluctuation in the ``critical regime'' of the 
oriented digital boiling model~\cite{GTW2001}. Recently the distribution
in one time case ($m=1$) has been also given in~\cite{BO2006c}. 
Our formula above gives a generalization to the multi-time version. 

In particular, the limiting distribution in the case $m=1$ can be 
described as
\begin{align}
\lim_{M\rightarrow\infty}\text{Prob}
\left(L(t,M)\ge\ell\right)
=
\sum_{k=0}^{\ell}\frac{(-1)^k}{k!}\sum_{x_1=0}^{\ell-1}
\cdots\sum_{x_k=0}^{\ell-1}
\det\left(\K_1(\t,x_l;\t,x_{l'})\right)_{l,l'=1}^k,
\label{r1m1}
\end{align}
where 
\begin{equation}
\K_1(\t,x_1;\t,x_2)=\sum_{m=0}^{x_2}
\psi_1(x_1-m,\t)\psi_2(x_2-m,\t).
\label{dHermite}
\end{equation}
Note that in~\eqref{r1m1} the summation on $k$ where $k\ge \ell+1$ can 
be omitted. Thus in this case, we can obtain exactly the limiting 
probability by calculating the determinants with finite rank. 
Another expression of the kernel~\eqref{dHermite} has recently 
appeared as the ``discrete Hermite kernel'' in~\cite{BO2006c}.
In~\cite{GTW2001}, the specific values of the probability are given in 
the case $\t=0$. 

\subsubsection{\bf Region 2 ($1/(1-q)< u <u_c$)}
This is the region where the effect of the defect particles does not 
affect the time evolution of the tagged particle and the dynamics of the 
ordinary TASEP is dominant. The typical picture of the time evolution of 
the whole particles from 1st to $M$th in this region is illustrated in 
Fig.~2(b). In this figure, we are interested in the position fluctuations 
of the bottommost particle.

Let us scale as
\begin{align}
\label{22scaling}
&t_j=uM + C(u) M^{\frac23}\t_j,
\\
&\ell_j=A_2(u_j)M-D(u)M^{\frac13}s_i,
\label{213}
\end{align}
where $1/(1-q) < u < u_c$ and $u_j=t_j/M$. 
$A_2(u)$ is defined in~\eqref{A2} and
\begin{align}
&C(u)=2(u-1)^{\frac56}
\left(1+\sqrt{\frac{1-q}{q(u-1)}}\right)^{\frac13}
\left(\sqrt{u-1}-\sqrt{\frac{q}{1-q}}\right)^{\frac13},\\
&D(u)=(u-1)^{\frac16}q^{\frac12}(1-q)^{\frac12}
\left(1+\sqrt{\frac{1-q}{q(u-1)}}\right)^{\frac23}
\left(\sqrt{u-1}-\sqrt{\frac{q}{1-q}}\right)^{\frac23}.
\label{22scalingf}
\end{align}
Note that the scaling exponent 1/3 in~\eqref{213} also appears in
an anomalous diffusion of the tagged particle~\cite{vB1991,MB1991}.
The scaling exponents $2/3$ and $1/3$ in~\eqref{22scaling} 
and~\eqref{213} are characteristic of the one-dimensional KPZ 
universality class. They are expected to be universal for all models
belonging to the KPZ universality class. 

Under the scaling defined above, we get the following theorem for the 
scaling function. This function is also universal but may depend on 
various elements such as initial conditions and the effect of the defect
particles. For example, as we will show in Theorem 2-3, the scaling
function in the region 3 is different from that in region 2 although
we take the same scaling~\eqref{22scaling} and~\eqref{213} in both 
regions. Thus, the function provides more detailed viewpoint in the KPZ 
universality than the scaling exponents. The proof of this theorem will 
be given in Section~\ref{pot223}.

\vspace{3mm}
\noindent
{\bf Theorem 2-2}
{\it
\begin{equation}
\lim_{M\rightarrow\infty}{\rm Prob}
\left(L(t_1,M)\ge\ell_1,L(t_2,M)\ge\ell_2,\cdots,
L(t_m,M)\ge\ell_m\right)=\det\left(1+\K_{2}\G\right).
\end{equation}
Here the right hand side is the Fredholm determinant 
defined as
\begin{align}
\label{Freddef}
&~\det\left(1+\K_{2}\G\right)\notag\\
&= \sum_{k=0}^{\infty} \frac{1}{k!} 
 \sum_{n_1=1}^m \int_{-\infty}^{\infty} d\xi_1 \cdots 
 \sum_{n_k=1}^m \int_{-\infty}^{\infty} d\xi_k 
 ~\G(\t_{n_1},\xi_1) \cdots \G(\t_{n_k},\xi_k)  
 \det(\K_2(\t_{n_l},\xi_l; \t_{n_{l'}},\xi_{l'}))_{l,l'=1}^k
\end{align}
where $\G(\t_j,\xi)~(j=1,\cdots,m)$ are defined in terms of
$\chi_{(a,b)}(x)$~\eqref{chi}, as
\begin{equation}
\G(\t_j,\xi)=-\chi_{(s_j,\infty)}(\xi)~(j=1,\cdots,m),
\label{defG}
\end{equation}
and the kernel $\K_2(\t_1,\xi_1;\t_2,\xi_2)$ is given by
\begin{equation}
 \label{K2def}
 \K_2(\t_1,\xi_1;\t_2,\xi_2)
 =
 \begin{cases}
  \int_0^{\infty} d\lambda e^{-\lambda(\t_1-\t_2)} \Ai(\xi_1+\lambda) \Ai(\xi_2+\lambda), 
  & \t_1 \geq \t_2 ,\\
  -\int_{-\infty}^0 d\lambda e^{-\lambda(\t_1-\t_2)} \Ai(\xi_1+\lambda) \Ai(\xi_2+\lambda), 
  & \t_1 < \t_2 .
 \end{cases}
\end{equation}
}
\vspace{3mm}

The kernel $\K_2$ is called the extended Airy kernel~\cite{FNH1999,
Ma1994}. The process characterized by the Fredholm determinant with 
this kernel is called the Airy process~\cite{Jo2003,PS2002b}. This 
process appears as the limiting process of the largest eigenvalue in 
Dyson's Brownian motion model~\cite{Dy1962} of the unitary class. The 
model is described as $N\times N$ Hermitian matrix where each 
independent element of $H$ obeys the Ornstein-Uhlenbeck process.
The transition probability density $\text{P}(H_i;H_j;\t)$ from matrix
$H_i$ to $H_j$ during $\t$ is given by
\begin{equation}
\text{P}\left(H_i;H_j;\t\right)
=Z_{\t}\exp\left(\frac{-\text{tr}
\left\{H_{j}-e^{-\t}H_i\right\}^2}
{1-e^{-2\t}}\right),
\end{equation}
where $Z_{\t}$ is the normalization constant. If we choose the initial 
matrix $H_0$ to be GUE random matrix, the joint density function of 
the probability that matrix $H_j$ is at time $t_j$ is represented as 
the Hermitian multi-matrix model,
\begin{align}
&~\text{P}\left(H_1,t_1;\cdots ; H_m,t_m\right)\notag\\
&=Z
 \prod_{j=1}^{m}\exp\left(\frac{-\text{tr}
\left\{H_{j}-e^{t_{j-1}-t_{j}}H_{j-1}\right\}^2}{1-e^{2(t_{j-1}-t_{j})}}\right)
\exp(-\text{tr}H_0^2),
\label{jointdyson}
\end{align}
where $Z$ is the normalization constant. Let $l^{(i)}$ be the largest 
eigenvalue of $H_i$ and we consider the quantity,
\begin{equation}
\label{laei}
\text{Prob}\left(l^{(1)}\le a_1,\cdots, l^{(m)}\le a_m \right).
\end{equation}
When we set
\begin{equation}
\label{mmset}
t_i=\frac{\t_i}{N^{\frac13}},~a_i=\sqrt{2N}
+\frac{s_i}{\sqrt{2}N^{\frac16}},
\end{equation}
the limiting distribution~\eqref{laei} is described as 
\begin{equation}
\lim_{N\rightarrow\infty}\text{Prob}\left(l^{(1)}\le a_1,\cdots, 
l^{(m)}\le a_m \right)=\det\left(1+\K_2\G\right).
\end{equation}
In the case $m=1$, the distribution is the GUE Tracy-Widom 
distribution~\cite{TW1994}.

Hence our Theorem 2-2 says that in the appropriate scaling limit the 
dynamics of the tagged particle in this region is equivalent to the 
dynamics of the largest eigenvalue of Dyson's Brownian motion of the 
unitary type. In the context of the one dimensional KPZ universality 
class, the Airy process has already appeared in the study of the 
polynuclear growth (PNG) model. In~\cite{Jo2003,PS2002b}, the process
has first appeared as the equal-time multi-point height 
fluctuation. Recently in~\cite{BO2006a}, it has been shown that the 
process also describes the correlations among any "space-like" points 
in the space-time plane. Our result is closely related to the latter
situation.

\subsubsection{\bf Region 3 ($u=u_c$)}\label{region3}
This region is the border between the region 2 where the normal 
particles in the TASEP are dominant, and the region 4 where only the 
finite number of defect particles are dominant. 

First we consider the case where the stay rates $\bar{q}_j~(j=1,2,
\cdots ,n)$ of the defect particles are distinct. Taking the same 
scaling as~\eqref{22scaling}--\eqref{22scalingf} with $u=u_c$, we 
obtain the following theorem.

\vspace{3mm}
\noindent
{\bf Theorem 2-3}
{\it
\begin{equation}
\lim_{M\rightarrow\infty}{\rm Prob}
\left(L(t_1,M)\ge\ell_1,L(t_2,M)\ge\ell_2,\cdots,
L(t_m,M)\ge\ell_m\right)=\det\left(1+\K_{3}\G\right),
\label{r3th23}
\end{equation}
where the Fredholm determinant and $\G(\t_j,\xi)$ in the right hand 
side are defined in~\eqref{Freddef} and~\eqref{defG} respectively.
The kernel is given by
\begin{align}
\label{kernelK3}
 &\quad 
 \K_{3}(\t_1,\xi_1;\t_2,\xi_2) \notag\\
 &=
 \begin{cases}
  \K_2(\t_1,\xi_1;\t_2,\xi_2)
  +\Ai(\xi_2) \int_0^{\infty} d\lambda e^{\t_1 \lambda}\Ai(\xi_1-\lambda) ,
  & \t_1<0,    \\
  \K_2(\t_1,\xi_1;\t_2,\xi_2)
  -\Ai(\xi_2) \int_0^{\infty} d\lambda e^{-\t_1\lambda}\Ai(\xi_1+\lambda)   
  +\Ai(\xi_2) e^{-\frac{\t_1^3}{3}+\xi_1\t_2},
  & \t_1>0 ,
 \end{cases}
\end{align}
where $\K_2$ is the extended Airy kernel~\eqref{K2def}.
}
\vspace{3mm}
\noindent

This kernel has appeared in~\cite{IS2004,IS2005} in the study of the 
height fluctuation property of the PNG model with external sources. 
When $\t_1=\cdots=\t_m=0$, it has been known that the Fredholm
determinant of this kernel is described as the distribution of the 
larger of the largest eigenvalues in two independent GOEs, which is 
denoted as GOE$^2$~\cite{BR2000,Fo2000p}.  On the other hand, when 
$\t_1=\cdots=\t_m=-\infty$, the Fredholm determinant becomes the GUE 
Tracy-Widom distribution. Thus \eqref{r3th23} describes the 
transition of the largest eigenvalue distribution between GUE and
GOE$^{2}$.

Next we consider the situation where the hopping rates 
$\bar{q}_j~(j=1,2,\cdots,n)$ are more or less the same.  In addition to the 
scaling~\eqref{22scaling}--\eqref{22scalingf}, we set the hopping 
rate of the defect particles as
\begin{equation}
\bar{q}_i=\bar{q}-\bar{q}(1-\bar{q})\frac{\eta_i}{D(u)M^{\frac13}},
\label{r3bq}
\end{equation}
where $D(u)$ is defined in~\eqref{22scalingf}. Note that the parameters 
$\eta_i (\ge 0) (i=1,\cdots,n)$ characterize the inhomogeneity of the 
hopping rates and the case $\eta_1=\eta_2=\cdots=\eta_n=0$ corresponds 
to the situation where the hopping rates degenerate completely. 

Under these settings, we obtain the following theorem. 

\vspace{3mm}
\noindent
{\bf Theorem 2-3'}
{\it
\begin{equation}
\label{FredK3}
\lim_{M\rightarrow\infty}{\rm Prob}
\left(L(t_1,M)\ge\ell_1,L(t_2,M)\ge\ell_2,\cdots,
L(t_m,M)\ge\ell_m\right)=\det\left(1+\K'_{3}\G\right),
\end{equation}
where the Fredholm determinant and $\G(\t_j,\xi)$ in the right hand 
side are defined in~\eqref{Freddef} and~\eqref{defG} respectively and
the kernel is given by
\begin{align}
\K'_3(\t_1,\xi_1;\t_2,\xi_2)
=&\K_2(\t_1,\xi_1;\t_2,\xi_2)
+\sum_{j=1}^n\frac{1}{2\pi}\int_{\Gamma_3}dw_1
\exp\left(i\xi_1w_1
+\frac{iw_1^3}{3}\right)\prod_{k=1}^j
\frac{1}{\eta_k-\t_1+iw_1}\notag\\
&~\times
\frac{1}{2\pi}
\int_{-\infty}^{\infty}dw_2\exp\left(i\xi_2w_2
+\frac{iw_2^3}{3}\right)\prod_{k=1}^{j-1}
(\eta_k-\t_2+iw_2),
\label{K3}
\end{align}
where the contour $\Gamma_3$ runs from $-\infty$ to $\infty$ passing the 
down side of the points $i(\eta_k-\t_1)~(k=1,\cdots,n)$. 
}
\vspace{3mm}

In the case $m=1$, this Fredholm determinant has appeared 
in~\cite{BBAP2005,Ba2006,DF2006p}. Note that if we set $\eta_j=0$
for some $j$ and $\eta_k=\infty$ for $k\neq j$, we can realize the 
situation where the maximum of $q_i~(i=1,2,\cdots,n)$ is unique such 
as the former case. Indeed in the setting above, the kernel 
$\K'_3$~\eqref{K3} is reduced to $\K_3$~\eqref{kernelK3}. In that 
sense,  the theorem is the generalization of Theorem 2-3. Later we 
will give only the proof of the theorem in Section~\ref{pot223}.

The Fredholm determinant~\eqref{FredK3} also describes the limiting 
largest eigenvalue distribution in the following multi-matrix model. 
Let $\{H_i\}_{i=1,\cdots,m}$ be the $N\times N$ Hermitian matrices. 
Analogous to \eqref{jointdyson}, the joint density function of the
model is defined as 
\begin{align}
&~\text{P}\left(H_1,t_1;\cdots;H_m,t_m\right)\notag\\
&=Z_{V}\exp(-\text{tr} H_0^2)\prod_{j=1}^{m}\exp\left(\frac{-\text{tr}
\left\{H_{j}-e^{t_{j-1}-t_j}H_{j-1}\right\}^2}{1-e^{2(t_{j-1}-t_{j})}}
\right)
\exp(\text{tr} VH_m).
\label{jointmulti}
\end{align}
Here $Z_V$ is the normalization constant and $t_i~(i=0,\cdots,m)$ are 
parameters of the model such that $t_{i-1}-t_i$ gauges the strength of 
the connection between matrix $H_{i-1}$ and $H_i$. $V=\text{diag}(v_1,
v_2,\cdots,v_n,0,0,\cdots\cdots)$ represents the external source with 
rank $n$ of the model. Under the scalings~\eqref{mmset} and  
\begin{equation}
v_i=\sqrt{2N}\left(1-\frac{\eta_i}{N^{\frac13}}\right),
\end{equation}
the limiting joint distribution of the largest eigenvalues 
$l^{(i)}$ of $H_i~(i=1,\cdots,m)$ is described as the Fredholm 
determinant~\eqref{FredK3},
\begin{equation}
\lim_{N\rightarrow\infty}\text{Prob}\left(l^{(1)}\le a_1,\cdots, 
l^{(m)}\le a_m \right)=\det\left(1+\K'_3\G\right).
\label{r3source}
\end{equation}
In the case $n=1$, \eqref{r3source} was shown in~\cite{IS2005}.
We can also show this equation for the general $n$ case in a
similar manner.

\subsubsection{Region 4 $(u_c<u)$}
This is the region where the effect of the defect particles has become 
dominant for the dynamics of the tagged particle. 

As in the region 3, we first consider the case $\bar{q}_1\neq 
\bar{q}_2\neq\cdots\neq \bar{q}_n$. In this case, one can expect that 
the dynamics is effectively the same as that of the slowest defect 
particle with stay rate $\bar{q}=\max\left(\bar{q}_j\right)$. To see 
nontrivial correlations, times $t_j~(j=1,\cdots,m)$ should be 
macroscopically separated. We set
\begin{align}
&\ell_j=A_G(u_j)M-D_G(u_j)M^{\frac12}s_j,
\label{12scaling}
\end{align}
where $u_j=t_j/M$, $A_G(u_j)$ is given in~\eqref{AG}, and
\begin{align}
D_G(u_j)=
\frac{1-\bar{q}}{\bar{q}}
\left(\frac{2\bar{q}^3}{1-\bar{q}}(u_j-1)-\frac{2\bar{q}^3q(1-q)}{(\bar{q}-q)^2(1-\bar{q})}
\right)^{\frac12}.
\end{align}
Under this setting and introducing the parameter $\t_j$ such that 
\begin{equation}
e^{\t_j}=D_G(u_j),
\label{r4tau}
\end{equation}
we get the theorem as follows.

\vspace{3mm}
\noindent
{\bf Theorem 2-4}
{\it
\begin{equation}
\lim_{M\rightarrow\infty}{\rm Prob}
\left(L(t_1,M)\ge\ell_1,L(t_2,M)\ge\ell_2,\cdots,
L(t_m,M)\ge\ell_m\right)=\det\left(1+\K_{G}\G\right),
\end{equation}
where the Fredholm determinant and $\G(\t_j,\xi)$ are defined 
in~\eqref{Freddef} and~\eqref{defG} respectively and
\begin{equation}
\K_G(\t_1,\xi_1; \t_2,\xi_2)=
\begin{cases}
\frac{\exp\left(-\xi_1^2\right)}{\sqrt{\pi}}
-\frac{\exp\left(-\frac{(\xi_2-e^{\t_1-\t_2}\xi_1)^2}
{1-e^{2(\t_1-\t_2)}}\right)}{\sqrt{\pi(1-e^{2(\t_1-\t_2)})}},
&\text{for~} \t_1< \t_2,\\
\frac{\exp\left(-\xi_1^2\right)}{\sqrt{\pi}}
,& \text{for~} \t_1\ge \t_2.
\label{r4KG}
\end{cases}
\end{equation}
}
\vspace{3mm}

This kernel represents the propagation of a Brownian particle (a 
particle obeying the Ornstein-Uhlenbeck process). Actually the two 
point distribution is calculated as
\begin{align}
&~\lim_{M\rightarrow\infty}\text{Prob}
\left(L(t_1,M)\ge\ell_1,L(t_2,M)\ge\ell_2\right)\notag\\
&= 1-\int_{s_1}^{\infty}d \xi_1 \K_{n}(\t_1,\xi_1;\t_1,\xi_1)
    -\int_{s_2}^{\infty}d \xi_2 \K_{n}(\t_2,\xi_2;\t_2,\xi_2)
    \notag\\
    &\quad+\frac12\int_{s_1}^{\infty}d\xi_1\int_{s_2}^{\infty}d\xi_2
    \left|
      \begin{array}{@{\,}cc@{\,}}
         \K_{n}(\t_1,\xi_1;\t_1,\xi_1)& \K_{n}(\t_1,\xi_1;\t_2,\xi_2) \\
         \K_{n}(\t_2,\xi_2;\t_1,\xi_1)& \K_{n}(\t_2,\xi_2;\t_2,\xi_2)
      \end{array}
    \right| \notag\\
    &\quad+\frac12\int_{s_1}^{\infty}d\xi_1\int_{s_2}^{\infty}d\xi_2
    \left|
      \begin{array}{@{\,}cc@{\,}}
         \K_{n}(\t_2,\xi_2;\t_2,\xi_2)& \K_{n}(\t_2,\xi_2;\t_1,\xi_1) \\
         \K_{n}(\t_1,\xi_1;\t_2,\xi_2)& \K_{n}(\t_1,\xi_1;\t_1,\xi_1)
      \end{array}
    \right| \notag\\
   &=\int_{-\infty}^{s_1}d\xi_1\int_{-\infty}^{s_2}d\xi_2
  \frac{e^{\frac{-(\xi_2-e^{\t_1-\t_2}\xi_1)^2}{1-e^{2(\t_1-\t_2)}}}}
{\sqrt{\pi(1-e^{2(\t_1-\t_2)})}}
  \frac{e^{-\xi_1^2}}{\sqrt{\pi}}.
  \label{orbrown}
\end{align}

Next we consider the case $\bar{q}_1\sim\bar{q}_2\sim\cdots\sim\bar{q}_n$.
The typical picture of the time evolution of the particles is depicted
in Fig.~2(c). In this figure, it seems that four groups of which each top
particle is the defect particle are formed and each group behaves like 
one particle in the TASEP. Thus in the region, one can expect that the 
system is effectively the same as $n$ TASEP particles.

In addition to the scaling~\eqref{12scaling}--~\eqref{r4tau}, we also 
set 
\begin{equation}
\bar{q}_i=\bar{q}-\bar{q}(1-\bar{q})\frac{2\e_i}{M^{\frac12}}.
\label{r4bq}
\end{equation}
Note that the scaling exponent 1/2 in~\eqref{12scaling} is the same as 
that in the case of the central limit theorem. However the following 
theorem indicates that the limiting process of $L(t,M)$ depends on the 
number $n$ of the defect particles. Note that this is not the 
information which the scaling exponent can detect.

\vspace{3mm}
\noindent
{\bf Theorem 2-4'}
{\it
\begin{equation}
\label{thm24'}
\lim_{M\rightarrow\infty}{\rm Prob}
\left(L(t_1,M)\ge\ell_1,L(t_2,M)\ge\ell_2,\cdots,
L(t_m,M)\ge\ell_m\right)=\det\left(1+\K^{(n)}\G\right).
\end{equation}
The Fredholm determinant and $\G(\t_j,\xi)$ in the right hand side are 
defined in \eqref{Freddef} and~\eqref{defG} respectively and the kernel 
is given by
\begin{equation}
 \K^{(n)}(\t_1, x_1;\t_2, x_2) 
 =
\tilde{K}^{(n)}(\t_1,x_1;\t_2,x_2)-\phi(\t_1,x_1; \t_2,x_2),
\label{kn}
\end{equation}
where
\begin{align}
&\tilde{K}^{(n)}(\t_1,\xi_1;\t_2,\xi_2)
=\frac{2}{(2\pi i)^2}\int_{\Gamma}dw_1\int_{\g}dw_2
\frac{\exp(w_2^2-w_1^2-2w_2\xi_2+2w_1\xi_1)}{e^{\t_1-\t_2}w_2-w_1}
\prod_{j=1}^n\frac{e^{-\t_2}w_2+\e_j}{e^{-\t_1}w_1+\e_j},
\label{tilk}
\\
&\phi(\t_1,\xi_1; \t_2,\xi_2)=
\begin{cases}
\frac{1}{\sqrt{\pi(1-e^{2(\t_1-\t_2)})}}
\exp\left(-\frac{(x_2-e^{\t_1-\t_2}\xi_1)^2}{1-e^{2(\t_1-\t_2)}}\right),
& {\text{for~}}\t_1<\t_2,\\
0,& {\text{for~}} \t_1 \ge \t_2.
\end{cases}
\label{npsi}
\end{align}
In \eqref{tilk}, $\Gamma$ represents the contour enclosing $-e^{\t_1}
\e_i~(i=1,\cdots,n)$ anticlockwise and $\g$ represents the arbitrary 
path running from $-i\infty$ to $i\infty$. 
}
\vspace{3mm}

The kernel above has been given in~\cite{IS2005,EM1999}. As in Theorem 
2-3', it is reduced to $\K_G$~\eqref{r4KG} if 
we set $\e=0$ for some $j$ and $\e_k=\infty$ for $k\neq j$. Thus, later we 
will give only the proof of this theorem in Section~\ref{pot24}.

The Fredholm determinant also describes the joint distribution function 
of the largest eigenvalue in the Hermitian multi-matrix 
model~\eqref{jointmulti}, where we assume the rank of 
$H_j~(j=1,\cdots,m)$ is $n$ and $V=\text{diag}(\e_1,\e_2,\cdots,\e_n)$. 
The joint distribution of the largest eigenvalues $l^{(i)}$ of $H_i$
can be described as the Fredholm determinant in the above theorem,
\begin{equation}
\text{Prob}\left(l^{(1)}\le s_1,\cdots, 
l^{(m)}\le s_m \right)=\det\left(1+\K^{(n)}\G\right).
\end{equation}
Note that in the equation above, we do not take the limit as the
rank $n$ goes to infinity.

\subsection{Scaling limit 2 ($M$ fixed)}\label{sclim2}

Here we consider another scaling limit such that $M$ is fixed and 
$t$ goes to infinity. We set
\begin{align}
\label{r5t}
&t_i=e^{2\t_i}T,\\ 
&\ell_i=(1-q)t_i-s_i\sqrt{2q(1-q)t_i}.
\label{r5elli}
\\
&q_i=q-\sqrt{\frac{2q(1-q)}{T}}\e_i,
\label{r5qi}
\end{align}
and consider the asymptotic behavior as $T$ goes to infinity. 
We get the following theorem. The proof of the theorem will be given 
in Section~\ref{pot3}.

\vspace{3mm}
\noindent
{\bf Theorem 3}
{\it
\begin{equation}
\lim_{T\rightarrow\infty}{\rm Prob}
\left(L(t_1,M)\ge\ell_1,L(t_2,M)\ge\ell_2,\cdots,
L(t_m,M)\ge\ell_m\right)=\det\left(1+\K^{(M)}\G\right),
\end{equation}
where the right hand side is the same as that of~\eqref{thm24'}
with $n=M$.
}
\vspace{3mm}

\subsection{Continuous limit}
In this subsection, we consider the continuous time version of the TASEP 
which is also usually discussed in the study of the ASEP. In the case 
without defect particles, the rule is defined as follows. Let 
$\tilde{t}$ be the time variable which can take any positive real 
number. Between time $\tilde{t}$ and $\tilde{t}+d\tilde{t}$, a particle 
can hop to its right neighboring site with probability $d\tilde{t}$ 
if the site is empty. If the site is occupied, the particle stays at 
the same site with probability 1.

We easily find that the continuous time version can be realized from 
the discrete time version defined in Section~\ref{model} by taking the 
limit, 
\begin{equation}
 1-q\rightarrow 0,~t\rightarrow\infty,
\label{cont1}
\end{equation}
with $\tilde{t}=(1-q)t$ fixed. In addition, the scaling limit in 
Section~\ref{sclim1} can also be studied by taking the 
$\tilde{t}\to\infty, M\to\infty$ limit with their ratio 
\begin{equation}
\tilde{u}=\tilde{t}/M,
\label{cont2}
\end{equation}
fixed. Applying~\eqref{cont1} and~\eqref{cont2} to the Theorems in 
Section~\ref{sclim1}, we can rewrite the theorems for the continuous 
time TASEP. Here we show only the result which corresponds to Theorem 
2-2. Analogous to~\eqref{22scaling} and~\eqref{213}, we scale time 
$\tilde{t}_j$ and the position $\ell_j$ as 
\begin{align}
\label{c22scaling}
&\tilde{t}_j=\tilde{u}M + C^{(c)}(\tilde{u}) M^{\frac23}\t_j,
\\
&\ell_j=A^{(c)}_2(\tilde{u}_j)M-D^{(c)}(\tilde{u})M^{\frac13}s_i,
\label{c213}
\end{align}
where $1<\tilde{u}$ and
\begin{align}
&A^{(c)}_2(\tilde{u})=
\left(\sqrt{\tilde{u}}-1\right)^2,\\
&C^{(c)}(\tilde{u})=2\tilde{u}^{\frac{5}{6}}\left(\sqrt{\tilde{u}}-1
\right)^{\frac13},\\
&D^{(c)}(\tilde{u})=\tilde{u}^{\frac16}\left(\sqrt{\tilde{u}}-1
\right)^{\frac23}.
\end{align}
Then we have
\begin{equation}
\lim_{M\rightarrow\infty}\text{Prob}
\left(L(\tilde{t}_1,M)\ge\ell_1,L(\tilde{t}_2,M)\ge\ell_2,\cdots,
L(\tilde{t}_m,M)\ge\ell_m\right)=\det\left(1+\K_{2}\G\right).
\end{equation}
Here the right hand side is defined in~\eqref{Freddef}--~\eqref{K2def}.

\section{From the TASEP to the Schur process}\label{ftts}
In this section, we discuss a relationship between the time evolution 
of a tagged particle in the TASEP with the step initial condition 
described in Section~\ref{model} and the stochastic growth of a Young 
diagram~\cite{OC2003a,OC2003b,BO2006a,BO2006b} which turns out to be 
a special case of the Schur process~\cite{OkRe2003,BoRa2006}. In the 
last part of this section, the proof of Theorem 1 in 
Section~\ref{multitimed} is obtained.

In the study of the TASEP, the technique using the enumeration of Young
diagrams was used in~\cite{Jo2000} to study the current fluctuation.
The discussion in~\cite{Jo2000} was based on the relation of the TASEP to 
the directed polymer problem on a matrix of which the elements obey 
the geometric distribution. In this section, on the other hand, we 
consider the correspondence between the directed polymer problem on a 
01 matrix~\cite{Se1998,J2001,GTW2001,GTW2002a} and the TASEP, which is a
similar but different mapping. For the analysis of the tagged particle 
problem, the mapping is more natural and convenient.

\subsection{TASEP and 01 matrix}\label{01table}
Let us consider an ensemble of $N\times M$ 01 matrices 
$\{a(i,j)\}_{i=1,\cdots,N,j=1,\cdots,M}$, whose matrix elements are 
either 0 or 1. All matrix elements are independent random variables and 
each element obeys the Bernoulli distribution,
\begin{align}
a(i,M+1-j)=
\begin{cases}
0, &\text{with probability~~} 1-q_j,\\
1, &\text{with probability~~} q_j.
\end{cases}
\label{01rule}
\end{align}
Here $1-q_j$ is taken to be the same value as the hopping rate of the 
$j$th particle of the TASEP defined in Section~\ref{model}. Each 
realization of the 01 matrix $a(i,j)$ can be translated to a time 
evolution of the TASEP as follows.
\begin{quote}
$a(i,M+1-j)=0$(resp. 1): Between time $i+j-2$ 
and $i+j-1$, the $j$th particle tries to hop 
to the right neighboring site (resp. stays at 
the same site).  When the right neighboring site 
is occupied at time $i+j-2$, particle can not move 
due to the exclusion effect even if $a(i,M+1-j)=0$.
\end{quote}
Since $a(i,M+1-j)=0$ could mean both hopping or stay depending on 
whether the target site is occupied or empty, the mapping from a $01$ 
matrix to the time evolution of the TASEP is not one-to-one. In Fig.~4, 
we depict an example of the time evolution in the TASEP and the 
corresponding two 01 matrices. Note that in the step initial 
condition, the $j$th particle cannot move until time $j-1$ and that 
the time evolution of the $j$th particle at time step between $i$ and 
$i+1$ does not depend on that of the $j-1$th particle at the time step. 
Thus one finds that the 01 matrix defined above has all informations 
needed for the time evolution of the $M$th particle from time 0 
through $M+N-1$.

Let us define a left-down path $\pi(N,M)$ of a $N\times M$ 01 matrix 
$\{a(i,j)\}_{i=1,\cdots,N,j=1,\cdots,M}$ by 
\begin{equation}
\pi(N,M)=\left\{ \{(i_k,j_k)\}_{1\le k}|a(i_k,j_k)=1, 
 1\le i_1< i_2<\cdots\le N, M\ge j_1\ge j_2\ge\cdots\ge 1
\right\}.
\label{drule}
\end{equation}
Note that the row indices $i_k~(k=1,\cdots)$ are {\it strictly} 
increasing while the column indices $j_k~(k=1,\cdots)$ are {\it weakly} 
decreasing. In $\pi$, the position of $(i_{k+1}, j_{k+1})$ is always 
located on the down side or left-down side of the position of 
$(i_k,j_k)$ in the 01 matrix. We also define the quantity $G(N,M)$ by
\begin{equation}
G(N,M)=\max_{\pi(N,M)}|\pi(N,M)|,
\label{defpi}
\end{equation}
where $|\pi|$ denotes the number of elements in the path $\pi$. 
If we regard the 01 matrix as the random 
media in a plane,  $\pi(N,M)$ as a spatial configuration of a polymer 
chain, and $|\pi(N,M)|$ as its energy, we can interpret~\eqref{defpi} 
as a statistical mechanical problem of a directed polymer. This kind 
of problem is studied in~\cite{Se1998,J2001,GTW2001,GTW2002a}.

Now we find that this maximum length of the left-down path is directly 
related to the position of a tagged particle as 
\begin{equation}
G(N,M) = d(N,M)
\label{prop1}
\end{equation}
where
\begin{equation}
d(N,M):=N-L(t=N+M-1,M)
\label{d}
\end{equation}
represents the number of times the $M$th particle stays at the same 
site from time 0 through $N+M-1$. The equation (\ref{prop1}) enables 
us to study the dynamics of the TASEP in terms of the 01 matrices and 
plays a fundamental role in our subsequent analysis. 

We can show~\eqref{prop1} in the following way. For the case $M=1$ 
(the case where we tag the first particle), we can easily check this 
equation, since $G(N,1)$ defined in~\eqref{defpi} is simply the total 
number of the value 1 in the sequence $\{a(k,1)\}_{k=1,\cdots,N}$ 
and it is clear that this quantity represents the number of times 
the particle stays from $t=0$ through $N$.

For the case $M\ge 2$, we can prove~\eqref{prop1} by mathematical 
induction about the row $N$. Here, we mainly consider only the case 
$M=2$. For the case $M\ge 3$, we can check~\eqref{prop1} in a similar 
fashion to this case.

When $N=1$, we can check~\eqref{prop1} easily for all four cases 
($a(1,1)=0,1$ and $a(1,2)=0,1$).

We assume that~\eqref{prop1} holds for $N=N_a$, and $M=1$ and $2$, i.e.,
\begin{equation}
d(N_a,1)=G(N_a,1),~~d(N_a,2)=G(N_a,2).
\label{assump}
\end{equation}
Under this assumption, we will show that $d(N_a+1,2)=G(N_a+1,2)$ 
as follows. From~\eqref{defpi}, one easily finds
\begin{equation}
G(N_a,2)\ge G(N_a,1),~~G(N_a,i)\ge G(N_b,i),
\label{pipro}
\end{equation}
where $1\le N_b\le N_a$, and $i=1$ and $2$. 
Considering these properties, we classify the problem into two cases,
\begin{description}
\item[(i)]
\begin{equation}
G(N_a,2)=G(N_a,1),~~\text{and~~}a(N_a+1,2)=1,
\label{1stcase}
\end{equation} 
\item[(ii)]
\begin{align}
\begin{cases}
G(N_a,2)> G(N_a,1) \text{~~and~~} a(N+1,2)=0\text{~or~} 1,\\
G(N_a,2)=G(N_a,1) \text{~~and~~} a(N+1,2)=0.
\end{cases}
\label{2ndcase}
\end{align}
\end{description}

The case (i) corresponds to the situation where the two particles occupy the
neighboring sites at $t=N_a+1$. Thus the second particle (the tagged 
particle) must stay at time step between  $N_a+1$ and $N_a+2$ . Hence 
one finds
\begin{equation}
d(N_a+1,2)=d(N_a,2)+1,
\label{da1}
\end{equation} 
regardless of the value of $a(N_a+1,1)$. On the other hand, due 
to~\eqref{defpi},~\eqref{pipro} and~\eqref{1stcase}, we have
\begin{equation}
G(N_a+1,2)=G(N_a,1)+a(N_a+1,2)=G(N_a,2)+1.
\label{pia1} 
\end{equation} 
Thus from~\eqref{assump},~\eqref{da1} and~\eqref{pia1}, we have for the 
case (i),
\begin{equation}
d(N_a+1,2)=G(N_a+1,2).
\label{induc}
\end{equation}

The case (ii) corresponds to the situation where at $t=N_a+1$, the 
distance between the first and the second particles is at least one 
site. Note that in this case we do not need to consider the exclusion 
effect of the first particle and thus the situation is essentially 
the same as the case $M=1$. We have
\begin{equation}
d(N_a+1,2)=d(N_a,2)+a(N_a+1,1).
\end{equation}
On the other hand, from~\eqref{pipro} and~\eqref{2ndcase} we can also 
check that
\begin{equation}
G(N_a+1,2)=G(N_a,2)+a(N_a+1,1).
\end{equation}
Thus we find that~\eqref{induc} holds also for the case (ii) and 
therefore~\eqref{prop1} holds for $M=2$.

In the case $M=k~(k\ge 3)$, we can also classify the situation into 
two cases as in (i) and (ii) for $M=2$, i.e., the case (i)' where all 
$k$ particles are packed and (ii)' where the particles form several 
groups which are separated by some successive empty sites. For the 
case (i)', we can check~\eqref{prop1} in a similar manner to the 
case $M=2$. For the case (ii)',  we can easily show that the case is 
essentially the same as the case (i)' for $M=m$ which is smaller 
than $k$.

\subsection{01 matrix and the dual Robinson-Schensted-Knuth 
correspondence}\label{01dRSK}
The quantity we would like to discuss is the multi-time distribution
of $L(t,M)$,
\begin{equation}
\text{Prob}_{\text{TASEP}}(L(t_1,M)\ge\ell_{1}, L(t_2,M)\ge\ell_{2},
\cdots, L(t_m,M)\ge\ell_{m}),
\end{equation}
where $\text{Prob}_{\text{TASEP}}$ represents the probability measure 
of the TASEP defined in Section~\ref{model}. Due to the result in 
Section~\ref{01table}(especially~\eqref{prop1} and~\eqref{d}), one 
finds
\begin{align}
&~\text{Prob}_{\text{TASEP}}(L(N_1+M-1,M)\ge\ell_{1},\notag\\
&\hspace{5cm}L(N_2+M-1,M)\ge \ell_{2},
\cdots,L(N_m+M-1,M)\ge\ell_m)\notag\\
&=\text{Prob}_{01}(G(N_1,M)\le N_1-\ell_{1},
~G(N_2,M)\le N_2-\ell_{2},\cdots,~G(N_m,M)\le N_m-\ell_m),
\label{t01t}
\end{align}
where $\text{Prob}_{01}$ represents the probability measure of 01 table 
defined in~\eqref{01rule} and $G(N_i,M)$ is defined in~\eqref{prop1}.

In this section, we further restate our problem in terms of the 
combinatorics of Young tableaux. For the definitions and basic 
properties of Young tableaux and related subjects, we refer the 
readers to \cite{St1999}. There is a bijective mapping between an 
$N\times M$ 01 matrix and a pair $(P,Q)$ where $P^{t}$ (the transpose 
of $P$) and $Q$ are the semistandard Young tableaux (SSYT) with the 
condition that the shape of $P$ is the same as that of $Q$. The 
mapping is called the dual Robinson-Schensted-Knuth (RSK) algorithm. 
In order to get $(P,Q)$, we first construct a two-line array (or 
generalized permutation),
\begin{equation}
\left(
\begin{array}{@{\,}cccccccccc@{\,}}
1&\cdots&1&2&\cdots&2&\cdots&N&\cdots&N\\
j^{(1)}_1&\cdots&j^{(1)}_{m_1}&j^{(2)}_{1}&\cdots
&j^{(2)}_{m_2}&\cdots&j^{(N)}_{1}&\cdots&j^{(N)}_{m_N}
\end{array}
\right),
\label{gperm}
\end{equation}
where $1\le j^{(i)}_1\le j^{(i)}_2\cdots\le M$, from the 01 matrix by 
listing $\binom{i}{j}$'s for which $a(i,j)=1$. Next, we construct $P$ 
(resp. $Q$) by arranging on a plane the figures in the second row 
(resp. the first row). Note that each positive integer constructing 
$P$(resp. $Q$) is less than $M$(resp. $N$). For details of the 
algorithm, see~\cite{St1999}. The generalized permutation and $(P,Q)$ 
corresponding to the 01 table in Fig.~5(b) are given in Fig.~6. 

Now we express $G(N,M)$~\eqref{defpi} as the quantity related to a pair 
$(P,Q)$. Let us denote the shape of $P$ (or $Q$) by the Young diagram, 
$\lambda(N,M)=(\lambda_1(N,M),\lambda_2(N,M),\cdots)$ where $\lambda_i\ge 0$ represents the 
length of $i$th row of the tableaux $P$ (or $Q$). In the example of 
Fig.~6(b), we see $\lambda=(4,3,2,2,2)$. Then we have 
\begin{equation}
G(N,M)=\lambda'_1(N,M).
\label{dl'}
\end{equation}
Here $\lambda'=(\lambda'_1,\lambda'_2,\dots)$ means the transpose of the Young diagram 
$\lambda$. Thus $\lambda'_1(N,M)$ is equal to the length of the first column of 
$\lambda(N,M)$.

Eq.~\eqref{dl'} may be shown by examining the dual RSK algorithm 
directly but is also understood as follows. For the sequence of 
the numbers $(i_1,i_2,\cdots)$ where $i\in (0,1,2,\cdots)$, we define
a nondecreasing (resp. nonincreasing) subsequence $(j_1,j_2,\cdots)$ such 
that $j_1\le j_2\le \cdots$ (resp. $j_1\ge j_2\ge \cdots$). 
From the construction it is not difficult to see that $G(N,M)$ is 
equivalent to the length of the longest nonincreasing subsequence of 
the second row $(j^{(1)}_{1},\cdots,j^{(1)}_{m_1},\cdots,j^{(N)}_{1},
\cdots, j^{(N)}_{m_N})$ in a generalized permutation~\eqref{gperm}.
Thus, $G(N,M)$ is also regarded as the length of the longest 
nondecreasing subsequence of the opposite sequence $(j^{(N)}_{m_N},
\cdots, j^{(N)}_{1},\cdots,j^{(1)}_{m_1},\cdots, j^{(1)}_{1})$ of the
second row in~\eqref{gperm}. We can apply the (normal) RSK 
algorithm~\cite{St1999} to this reverse sequence to obtain a SSYT. 
Let us denote the SSYT by $R$. Then (\ref{dl'}) is a consequence of 
the facts that the length of the longest nondecreasing sequence in the 
reverse sequence is equal to the length of the first row of $R$ (this is 
known as a property of the normal RSK algorithm~\cite{Jo2000,BR2001c}) 
and the symmetry property, 
\begin{equation}
P^{t}=R,
\label{ptr}
\end{equation}
whose proof can be found in Appendix A of~\cite{Ful1999}. Notice that 
in the example of Fig.~5(a), $L(9,4)=1$ thus $d(6,4)=5$ while in 
Fig.~6(b), $\lambda'_1=5$. 

Thus from~\eqref{dl'}, we get  
\begin{align}
&~\text{Prob}_{01}(G(N_1,M)\le N_1-\ell_{1},
~G(N_2,M)\le N_2-\ell_{2},\cdots,~G(N_m,M)\le N_m-\ell_m)\notag\\
&=\text{Prob}_{\lambda}(\lambda'_1(N_1,M)\le N_1-\ell_{1},~\lambda'_1(N_2,M)
\le N_2-\ell_{2},
\cdots,\lambda'_1(N_m,M)\le N_m-\ell_{m}).
\label{01y}
\end{align}
Here $\text{Prob}_{\lambda}$ in the right hand side represents the 
probability measure of the set of Young diagrams obtained by the dual 
RSK algorithm from the 01 matrix defined in \eqref{01rule} and 
$\lambda(N_i,M)$ means the Young diagram obtained by applying the dual RSK 
algorithm to the $N_i\times M$ 01 submatrix 
$\{a(i,j)\}_{1\le i\le N_i, 1\le j\le M}$ of the $N\times M$ 01 
matrix~\eqref{01rule}.

From \eqref{t01t} and~\eqref{01y}, we managed to interpret the time 
evolution of a tagged particle in the TASEP as the time evolution of 
$\lambda'_1$, the length of the first column, regarding $N_i$ as a time 
parameter.

\subsection{Growth of Young diagrams and Schur process}
In order to investigate the right hand side of \eqref{01y}, we consider 
the following probability,
\begin{equation}
\text{Prob}_{\lambda}(\lambda(1,M)=\lambda^{(1)},~\lambda(2,M)=\lambda^{(2)},
\cdots,\lambda(N,M)=\lambda^{(N)}=\lambda)
\label{goyoung}
\end{equation}
for a given set of Young diagrams, $\{\lambda^{(k)}\}_{k=1,\ldots,N}$. 
This equation describes the growth of a Young diagram. 
Eq.~\eqref{goyoung} represents the joint distribution of them. Fig.~7 
illustrates the growth of Young diagram which corresponds to Fig.~5(b).
Notice that from~\eqref{prop1} and~\eqref{dl'}, the growth of 
$\lambda'_1(i,M)$ describes the time evolution of $M$th particle in the 
TASEP. Let us denote the $Q^{(k)}$ as the part of the $Q$ tableau 
obtained by picking up only the symbols $1,2,\cdots ,i$. From the rule of the 
dual RSK algorithm~\cite{St1999}, we find that $Q^{(k)}$ is equivalent 
to the $Q$ tableau obtained by applying the dual RSK algorithm to the 
the submatrix $\{a(i,j)_{i=1,\ldots,k,j=1,\ldots,M}\}$. Thus one has
\begin{equation}
\text{sh}(Q^{(i)})=\lambda(i,M),
\label{Ql}
\end{equation}
where $\text{sh}(Q^{(i)})$ denotes the shape of $Q^{(i)}$. Hence the 
$Q$ tableau records the growth of Young diagram, that is, it one-to-one 
corresponds to the set of the Young diagrams 
$\{\lambda(i,M)\}_{i=1,\cdots,N}$. 
Compare the tableau $Q$ in Fig.~6(b) with $\{\lambda(i,M)\}_{i=1,\cdots,6}$ 
in Fig.~7. We can recognize that $\text{sh}(Q^{(i)})$ constructed from 
the tableau $Q$ in Fig.~6(b) is equivalent to $\lambda(i,M)$ in Fig.~7.
Hence we have
\begin{align}
&~\text{Prob}_{\lambda}(\lambda(1,M)=\lambda^{(1)},~\lambda(2,M)=\lambda^{(2)},
\cdots,\lambda(N,M)=\lambda^{(N)})\notag\\
&=
\text{Prob}_{\{P,Q\}}(\text{sh}(Q^{(k)})=\lambda^{(k)},k=1,\ldots,N).
\label{grsk}
\end{align}
Here $\text{Prob}_{\{P,Q\}}$ represents the probability measure on the 
set of the pairs $\{P,Q\}$ obtained by the dual RSK algorithm from the 
$N\times M$ 01 matrices defined in \eqref{01rule}. Hence it is found 
that the growth process of the Young diagram is characterized by the 
$Q$ tableau in the dual RSK correspondence. This type of growth process 
has been also discussed in~\cite{OC2003a,OC2003b,BO2006a,BO2006b}.

Remembering $P$ is constructed from the second row of a generalized 
permutation, we find that the symbol $j$ in $P$ represents the column 
index of the element of the 01 matrix where $a(i,j)$=1. Thus the total 
number of $j$s in $P$ is equal to the total number of the figure 1 in 
the $j$th column of the 01 matrix. Hence, for a given $P'$, one finds
\begin{align}
&~\text{Prob}_{\{P,Q\}}
(P=P',\text{sh}(Q^{(k)})= \lambda^{(k)}(k=1,\cdots,N))\notag\\
&=
\begin{cases}
q_M^{\sh1(P')}q_{M-1}^{\sh2(P')}\cdots q_1^{\sh M(P')}
\times (1-q_M)^{N-\sh1(P')}(1-q_{M-1})^{N-\sh2(P')}
\cdots (1-q_1)^{N-\sh M(P')},\\ 
{\text{if $\lambda^{(1)}$ and $\lambda^{(i+1)}/\lambda^{(i)}~(i=1,\cdots,N-1)$ 
have no two squares in the same column,}}
\\
0,~\text{otherwise.}
\end{cases}
\label{ssyt}
\end{align}
where $\sh i(P')$ means the total number of the symbol $i$ in $P'$. 
Applying the above equation to the combinatorial definition of the 
Schur function,
\begin{equation}
s_{\lambda/\mu}(a_1,\cdots,a_N)
=\sum_{S}a_1^{\sh1(S)}
\cdots a_N^{\sh N(S)},
\label{cdefschur}
\end{equation}
where the summation is taken to all SSYT $S$ with shape $\lambda/\mu$, we 
have 
\begin{align}
&~\text{Prob}_{\{P,Q\}}(\text{sh}(Q^{(k)})=\lambda^{(k)},k=1,\ldots,N)
\notag\\
&=\sum_{P'}\text{Prob}_{\{P,Q\}}
(P=P',\text{sh}(Q^{(k)})=\lambda^{(k)},k=1,\ldots,N)\notag\\
&=
(1-q_1)^N\cdots(1-q_M)^Ns_{\lambda'^{(N)}}(p_M,\cdots,p_1) \notag\\
&\quad
~s_{\lambda^{(1)}}(1,0,\cdots)s_{\lambda^{(2)}/\lambda^{(1)}}(1,0,\cdots)
\times\cdots\times s_{\lambda^{(N)}/\lambda^{(N-1)}}(1,0,\cdots),
\label{sp}
\end{align}
where $p_i=q_i/({1-q_i})$. The factor $s_{\lambda^{(1)}}(1,0,\cdots)
\cdots$ ensures that the probability is zero unless $\lambda^{(1)}$ and 
$\lambda^{(i+1)}/\lambda^{(i)}~(i=1,\cdots,N-1)$ have no two squares in the same 
column. Thus combining ~\eqref{grsk} with~\eqref{sp}, we finally find
\begin{align}
&~\text{Prob}_{\lambda}(\lambda(1,M)=\lambda^{(1)},~\lambda(2,M)=\lambda^{(2)},
\cdots,\lambda(N,M)=\lambda^{(N)})\notag\\
&=(1-q_1)^N\cdots(1-q_M)^Ns_{\lambda'^{(N)}}(p_M,\cdots,p_1)\times
\notag\\
&~s_{\lambda^{(1)}}(1,0,\cdots)s_{\lambda^{(2)}/\lambda^{(1)}}(1,0,\cdots)
\times\cdots\times s_{\lambda^{(N)}/\lambda^{(N-1)}}(1,0,\cdots).
\label{taschur}
\end{align}
This measure is the special case of the joint distribution function in 
the Schur process~\cite{OkRe2003,BoRa2006}. Since the Schur function is 
expressed as a determinant, the function~\eqref{taschur} is described 
as a product of determinants. Furthermore, it is known that the process 
has a remarkable mathematical structure that the correlation function of 
$\{\lambda_i^{(j)}\}_{i=1,2,\cdots,~j=1,\cdots N}$ can be expressed as a 
determinant and its scaling limit can be discussed exactly. The Schur 
process also appears in other fields of physics and mathematics such as 
random growth process~\cite{Jo2003}, melting problem of a 
three-dimensional crystal~\cite{Fe2003}, 
random tiling model~\cite{Jo2005}, and so on. 

\subsection{Proof of Theorem 1}
From~\eqref{t01t},\eqref{01y} and \eqref{taschur}, we find
\begin{align}
&~\text{Prob}_{\text{TASEP}}(L(t_1,M)\ge \ell_{1}, L(t_2,M)\ge \ell_{2},
\cdots, L(t_m,M)\ge \ell_{m}) \notag\\
&=\text{Prob}_{\text{SP}}(\lambda'_1(N_1,M)\le N_1-\ell_{1},~\lambda'_1(N_2,M)
\le N_2-\ell_{2},
\cdots,\lambda'_1(N_m,M)\le N_m-\ell_{m}).
\label{sp2}
\end{align}
Here $\text{Prob}_{\text{SP}}$ represents the probabilistic measure of 
the Schur process~\eqref{sp}.

In general, the Schur process is defined as follows. For the set of the 
Young diagrams such that
\begin{equation}
\lambda^{(1)}\supset\mu^{(1)}\subset\lambda^{(2)}\supset\mu^{(2)}
\cdots\supset\mu^{(N-1)}\subset\lambda^{(N)},
\label{spdef}
\end{equation}
the following weight is assigned
\begin{equation}
s_{\lambda^{(1)}}(\rho_0^+)
s_{\lambda^{(1)}/\mu^{(1)}}(\rho_1^-)s_{\lambda^{(2)}/\mu^{(1)}}(\rho_1^+)
\cdots
s_{\lambda^{(N)}/\mu^{(N-1)}}(\rho_{N-1}^+)s_{\lambda^{(N)}}(\rho_N^-).
\end{equation}
Here $s_{\rho/\mu}(\rho)$ is the Schur function with the specialization 
of algebra $\rho$. In our case, we notice only two cases
\begin{equation}
s_{\lambda/\mu}(\rho)=
\begin{cases}
s_{\lambda/\mu}(a_1,\cdots,a_N), &{\text{for~}}
\rho=\rho(z)=\prod_{j=1}^N\frac{1}{1-a_jz},\\
s_{\lambda'/\mu'}(a_1,\cdots,a_N), &{\text{for~}}
\rho=\rho(z)=\prod_{j=1}^N(1+a_jz),
\end{cases}
\end{equation}
where $s_{\lambda/\mu}(a_1,\cdots,a_N)$ is defined in~\eqref{cdefschur}.
Substituting to~\eqref{spdef},
\begin{align}
&\rho_i^-(z)=1+z,~~(1\le i\le N-1),\notag\\
&\rho_{N}^-=\prod_{i=1}^M\frac{1}{1-p_i z},\notag\\
&\rho_j^+(z)=1,~~(0\le j\le N-1),
\end{align}
where $p_i={q_i}/{(1-q_i)}$ and applying Theorem 1 in~\cite{OkRe2003} 
or Theorem 2.2 in~\cite{BoRa2006} to this case, we finally obtain the 
Fredholm determinant representation(\eqref{fFreddef}-\eqref{thm1f}) 
of \eqref{sp2}.

\section{Asymptotics}\label{asymp}
In this section, we discuss the scaling limit of the multi-time 
distribution function~\eqref{1} by applying the saddle point method 
to the kernel~\eqref{kernel}-\eqref{thm1f} in Theorem 1. 

We consider two types of scaling limit.  The first case is explained 
in Section~\ref{sclim1}.  We focus on the situation where both time 
$t$ and the label of the tagged particle $M$ tend to infinity when there 
are $n$ defect particles with the stay rates $\{\bar{q}_j\}_{j=1,
\cdots,n}$ in front of the tagged particle. The results on the limiting 
distributions are summarized as Theorem 2. Their proofs are 
given in Section 5.1, 5.2 and 5.3.

In the second case, we take $t$ to be infinite with $M$ fixed. The 
limiting distribution is described in Theorem 3 in 
Section~\ref{sclim2}, whose proof is given in Section 5.4.

\subsection{Region 1 (proof of Theorem 2-1)}\label{pot21}
In this section, we consider the scaling limit in the region 1 in 
Fig.~3. We scale time $t_i$ as~\eqref{r1ti}. In order to calculate the 
limiting distribution, we rewrite the 
kernel~\eqref{kernel}-\eqref{thm1f} with the condition~\eqref{1dp} as 
\begin{align}
&~K(t_1,x_1;t_2,x_2)\notag\\
&=\tilde{K}(t_1,x_1;t_2,x_2)-\phi_{t_1,t_2}(x_1,x_2)\notag\\
&=\frac{1}{(2\pi i)^2}\int_{C_{R'_1}}\frac{dz_1}{z_1}
\int_{C_{R'_2}}\frac{dz_2}{z_2}\frac{z_1}{z_1-z_2}
\frac{(1+1/z_2)^{t_2-M+1}}{(1+1/z_1)^{t_1-M+1}}
\left(\frac{1-pz_2}{1-pz_1}\right)^{M-n}
\prod_{i=1}^n\frac{1-\bar{p}_iz_2}{1-\bar{p}_iz_1}
\frac{z_2^{x_2}}{z_1^{x_1}},
\label{r1kernel}
\end{align}
where $p=q/(1-q)$ and 
$\bar{p}_i=\bar{q}_i/(1-\bar{q}_i)$. $C_R$ denotes the contour with 
radius $R$ enclosing the origin anticlockwise and 
$R'_{i}~(i=1,2)$ satisfy the condition,
$1<R'_2<R'_1<1/\bar{p}_i$ for $t_1\ge t_2$ and
$1<R'_1<R'_2<1/\bar{p}_i$ for $t_1<t_2$.
Using~\eqref{r1kernel} and the relation
\begin{equation}
\frac{z_1}{z_1-z_2}=
\begin{cases}
\sum_{m=0}^{\infty}\left(\frac{z_2}{z_1}\right)^m,&
t_1\ge t_2,\\
-\sum_{m=0}^{\infty}\left(\frac{z_1}{z_2}\right)^{m+1},&
t_1<t_2,
\end{cases}
\end{equation}
one finds 
\begin{equation}
K(t_1,x_1;t_2,x_2)=
\begin{cases}
\sum_{m=0}^{\infty}\Psi_1(x_1+m,t_1)\Psi_2(x_2+m,t_2),&
t_1\ge t_2,\\
-\sum_{m=0}^{\infty}\Psi_1(x_1-m-1,t_1)\Psi_2(x_2-m-1,t_2),
& t_1< t_2.
\end{cases}
\label{kernel1}
\end{equation}
Here the function $\Psi_1(m,t_1,x_1)$ and $\Psi_2(m,t_2,x_2)$ are 
given by
\begin{align}
&\Psi_1(x_1,t_1)
=\frac{1}{2\pi i}\int_{C_{R_1}} \frac{dz}{z} 
\frac{1}{(1+1/z)^{t_1-M+1}(1-pz)^{M}}
\prod_{i=1}^{n}\frac{1-pz}{1-\bar{p}_iz}
\frac{1}{z^{x_1}},
\label{r1Psi}
\\
&\Psi_2(x_2,t_2)
=\frac{1}{2\pi i}\int_{C_{1}} \frac{dw}{w} (1+1/w)^{t_2-M+1}
(1-pw)^{M}
\prod_{i=1}^{n}\frac{1-\bar{p}_iw}{1-pw}
w^{x_2},
\label{r1Psi2}
\end{align}
where the radius $R_1$ of the contour $C_{R_1}$ in~\eqref{r1Psi} is 
taken such that $1<R_1<1/\bar{p}_i$. In the following discussion, we
evaluate the asymptotics of the kernel~\eqref{r1kernel} by applying 
the saddle point method to the functions $\Psi_i(x,t)~(i=1,2)$.

First we consider the asymptotics of $\Psi_1(x_1,t_1)$. 
Substituting~\eqref{r1ti} with~\eqref{r1Psi}, we set
\begin{equation}
\Psi_1(x_1,t_1)
=\frac{1}{2\pi i}\int_{C_{R_1}} \frac{dz}{z} 
e^{-Mf(z)}\frac{1}{(1+z)^{1+M^{\frac12} D_1 \t_1}}
\prod_{i=1}^{n}\frac{1-pz}{1-\bar{p}_iz}
\frac{1}{z^{x_1-t_1+M-1}},
\label{r1Psi1'}
\end{equation}
where 
\begin{equation}
f(z)=p\log (1+z)+\log(1-pz).
\label{r1f}
\end{equation}
From the equation $f'(z_c)=0$, we find the critical point $z_c$ is 
given by
\begin{equation}
z_c=0.
\end{equation}
Scaling the variable $z$ around the critical point as
\begin{equation}
z=z_c+\frac{z'}{D_1  M^{\frac12}}=
\frac{z'}{D_1  M^{\frac12}},
\label{zscale}
\end{equation}
where $D_1=\sqrt{q}/(1-q)=\sqrt{p(1+p)}$, we evaluate the asymptotics of 
the term $e^{-Mf(z)}$ in~\eqref{r1Psi1'} by the saddle point method,
\begin{align}
e^{-Mf(z)}\sim e^{-M\left(f(0)+\frac{f''(0)}{2D_1^2M}z'^2\right)}
=e^{\frac{z'^2}{2}},
\end{align}
as $M\rightarrow\infty$. Next we consider the asymptotics of other terms 
in~\eqref{r1Psi1'}. From~\eqref{zscale}, one obtains
\begin{align}
\frac{1}{(1+z)^{M^{\frac12}D_1 \t_1+1}}
\sim e^{-\t_1z'},~\prod_{i=1}^n
\frac{1-pz}{1-\bar{p}_iz}\sim 1.
\end{align}
Transforming $x_i$ into $x'_{i}~(i=1,2)$, 
\begin{equation}
x_i=t_i-M+1-x'_i,
\label{xscale}
\end{equation}
and considering~\eqref{zscale}, one gets
\begin{equation}
\frac{1}{z^{x_1-t_1+M-1}}\sim
\frac{z^{x'_1}}{D_1^{x'_1}M^{\frac{x'_1}{2}}}.
\label{r1oterm2}
\end{equation}
Note that by the transformation~\eqref{xscale}, the function 
$g(t_i,x)$~\eqref{chi} in the Fredholm determinant~\eqref{fFreddef}
is changed to $g(\t_i,x')=-\chi_{(-\infty,\ell_i)}(x')$. 
From~\eqref{zscale}-\eqref{r1oterm2}, we obtain the asymptotic form 
of $\Psi_1(x_1,t_1)$,
\begin{equation}
\Psi_1(x_1,t_1)\sim
\frac{D_1^{-x'_1}M^{\frac{-x'_1}{2}}}{2\pi i}
\int_{-i\infty+\e}^{i\infty+\e}
dz'e^{\frac{z'^2}{2}-\t_1z'}{z'^{x'_1-1}},
\label{Psi1}
\end{equation}
where $\e>0$.

Next we consider the asymptotics of 
$\Psi_2(x_2,t_2)$~\eqref{r1Psi2}. Similar to $\Psi_1(x_1,t_1)$, we 
set
\begin{equation}
\Psi_2(x_2,t_2)
=\frac{1}{2\pi i}\int_{C_1}\frac{dw}{w} 
e^{Mf(w)}(1+z)^{1+M^{\frac12} D_1 \t_2}
\prod_{i=1}^{n}\frac{1-\bar{p}_iw}{1-pw}
w^{x_2-t+M-1}.
\end{equation}
Here $f(w)$ is defined in~\eqref{r1f} and $C_1$ denotes the contour 
enclosing the origin anticlockwise with radius 1. Applying the saddle 
point method to this equation in the same way as $\Psi_1(x_1,t_1)$,
we get
\begin{equation}
\Psi_2(x_2,t_2)\sim
\frac{D_1^{x'_2}M^{\frac{x'_2}{2}}}{2\pi i}\int_{C_1}
\frac{dw'}{w'^{x'_2+1}} e^{-\frac{w'^2}{2}+\t_2w'}.
\label{Psi2}
\end{equation}
Thus from~\eqref{kernel1},~\eqref{Psi1}, and~\eqref{Psi2}, we find
\begin{align}
&~\lim_{M\rightarrow\infty}K(t_1,x_1;t_2,x_2)\notag\\
&=
\begin{cases}
(D_1\sqrt{M})^{x'_2-x'_1}
\sum_{m=0}^{\infty}\psi_1(x'_1-m,\t_1)\psi_2(x'_2-m,\t_2),
& \t_1\ge \t_2,\\
-(D_1\sqrt{M})^{x'_2-x'_1}
\sum_{m=0}^{\infty}\psi_1(x'_1+m+1,\t_1)\psi_2(x'_2+m+1,\t_2),
& \t_1< \t_2,
\end{cases}
\end{align}
where
\begin{align}
&\psi_1(x'_1,\t_1)=\frac{1}{2\pi i}\int_{-i\infty+\e}^{i\infty+\e}
dz e^{\frac{z^2}{2}-\t_1z}z^{x'_1-1},\\
&\psi_2(x'_2,\t_2)=\frac{1}{2\pi i}\int_{C_1} \frac{dw}{w^{x'_2+1}} 
e^{-\frac{w^2}{2}+\t_2w}.
\end{align}
Introducing the function $D_{n}(z)$ by 
\begin{equation}
D_{x-1}(\t)=\sqrt{2\pi}e^{\frac{\t^2}{4}}\psi_1(x,\t),
\end{equation}
one finds that the function satisfies Weber's equation~\cite{AAR1999},
\begin{equation}
\frac{d^2D_{n}(z)}{dz^2}+\left(n+\frac12-\frac14z^2\right)
D_{n}(z)=0.
\label{weber}
\end{equation}
The initial condition was given in~\cite{GTW2001} as
\begin{align}
&D_{n}(0)=\frac{2^{\frac{n+1}{2}}}{\sqrt{2\pi}}
\sin\left(\frac{\pi}{2}(n+1)\right)
\Gamma\left(\frac{n+1}{2}\right),\\
&D'_{n}(0)=-D_{n+1}(0),
\label{pcf}
\end{align}
where $\Gamma(x)$ represents the Gamma function. The function 
satisfying~\eqref{weber}--\eqref{pcf} is known as the parabolic 
cylinder function~\cite{AAR1999}. In addition, one easily finds 
$\psi_2(x,\t)$ can be represented as~\eqref{th1-3} from the 
integral representation of the Hermite polynomial 
$H_n(x)$~\cite{AAR1999},
\begin{equation}
H_n(x)=\frac{n!}{2\pi i}\oint\frac{dz}{z^{n+1}}e^{2xz-z^2}.
\end{equation}

Hence, noticing the prefactor $(D_1\sqrt{M})^{x'_2-x'_1}$ does not 
contribute to the determinant in~\eqref{fFreddef}, we get the limiting 
kernel~\eqref{th1-1}-\eqref{th1-3}.

\subsection{Regions 2 and 3 (proof of Theorems 2-2 and 2-3')}
\label{pot223}
In this section we consider the scaling limit of the regions 2 and 3
in Fig.~3. We fix the scaled time $u$~\eqref{1sl} such that 
$1/(1-q)<u<u_c$ for the region 2 and $u=u_c$ for the region 3, where 
$u_c=(\bar{q}^2-2q\bar{q}+q)/(\bar{q}-q)^2$. We scale the time $t_j$, 
the position $\ell_j$ of the tagged particle at $t_j$, and the 
stay rates of the defect particles $\bar{q}_i~(i=1,\cdots,n)$ 
as~\eqref{22scaling},~\eqref{213} and~\eqref{r3bq} respectively.

First we analyze the asymptotics of $\tilde{K}(t_1,x_1;t_2,x_2)$ 
in~\eqref{Ktxtx} by the saddle point method. Using the variable 
$\mu(u)$ which will be fixed later, we set
\begin{align}
\tilde{K}(t_1,x_1;t_2,x_2)
&=\frac{1}{(2\pi i)^2}\int_{C_{R_1}}\frac{dz_1}{z_1}\int_{C_{R_2}}
\frac{dz_2}{z_2}
\exp\left(M(f_{u_2}(z_2)-f_{u_1}(z_1))\right)\frac{z_1}{z_1-z_2}\notag\\
&\times\frac{z_2^{x_2-\mu(u_2)M}}{z_1^{x_1-\mu(u_1)M}}
\left(\frac{1-pz_1}{1-pz_2}\right)^{n}
\prod_{i=1}^{n}\frac{1-\bar{p}_iz_2}{1-\bar{p}_iz_1}
\frac{1+1/z_2}{1+1/z_1}.
\label{kerneldint}
\end{align}
Here $u_i=t_i/M$, $p=q/(1-q)$, and $\bar{p}_i=\bar{q}_i/(1-\bar{q}_i)$. 
The function $f_{u}(z)$ is defined as
\begin{align}
f_{u}(z)=(u-1)\log(1+z)+\log(1-pz)+(\mu(u)-u+1)\log(z).
\label{fnr23}
\end{align}

We fix the value of $\mu(u)$ in such a way that two saddle points
of $f_{u}(z)$ merge to one point. We have
\begin{align}
\label{mur23}
\mu(u)=u-1-A_2(u)=\frac{p(u-2)+2\sqrt{p(u-1)}}{1+p},
\end{align}
and the double critical point $z_c(u)$ of $f_{u}(z)$is obtained as
\begin{equation}
z_c(u)=\frac{\sqrt{u-1}-\sqrt{p}}{\sqrt{p^2(u-1)}+\sqrt{p}}.
\label{dspr23}
\end{equation}
Note that 
\begin{equation}
f'_{u}(z_c(u))=f''_{u}(z_c(u))=0
\label{r23dcritical}
\end{equation}
is satisfied with $\mu(u)$ fixed as~\eqref{mur23}.

Thanks to the property that $z_c(u)\le 1/\bar{p}_i$ when $1/(1-q)<u\le 
u_c$ which is the condition of regions 2 and 3, it is found that we can 
deform the contours $C_{R_i}~(i=1,2)$ around the double saddle 
point~\eqref{dspr23}. Thus we scale $z_1$ and $z_2$ as
\begin{align}
z_1=z_c(u_1)\left(1-\frac{iw_1}{D(u)M^{\frac13}}\right),~
z_2=z_c(u_2)\left(1+\frac{iw_2}{D(u)M^{\frac13}}\right),
\label{zwscale}
\end{align}
where $D(u)$ is defined in~\eqref{22scalingf}. 
Noting~\eqref{r23dcritical} and the relation
\begin{equation}
-\frac{f'''(z_c(u))z_c^3(u)}{2}=D(u)^3,
\label{r23drelation}
\end{equation}
we have
\begin{align}
\exp\left(M(f_{u_2}(z_2)-f_{u_1}(z_1))\right)
&\sim\frac{\exp(Mf_{u_2}(z_c(u_2))+
M\frac{f_{u_2}^{'''}(z_c(u_2))}{6}
(z_2-z_c(u_2))^3)}{\exp(
Mf_{u_1}(z_c(u_1))+M\frac{f_{u_1}^{'''}(z_c(u_1))}{6}
(z_1-z_c(u_1))^3)}\notag\\
&\sim\exp\left(M(f_{u_2}(z_c(u_2))-f_{u_1}(z_c(u_1)))+
\frac{i(w_1^3+w_2^3)}{3}\right).
\end{align}
Hence we had the asymptotic form of the term 
$\exp\left(M(f_{u_2}(z_2)-f_{u_1}(z_1))\right)$
in~\eqref{kerneldint}. Next we consider that of other terms 
in~\eqref{kerneldint}. Combining~\eqref{zwscale} with the relation
\begin{equation}
C(u)\frac{z'_c(u)}{z_c(u)}=\frac{1}{D(u)M^{\frac13}},
\label{r23crelation}
\end{equation}
we find  
\begin{align}
z_1\sim z_c(u)\left(1+\frac{\t_1-iw_1}{D(u)M^{\frac13}}\right),~
z_2\sim z_c(u)\left(1+\frac{\t_2+iw_2}{D(u)M^{\frac13}}\right).
\label{zw0scale}
\end{align}
Furthermore, associated with~\eqref{213}, we scale $x_i~(i=1,2)$ as
\begin{align}
x_i=(u-1)M-A_2(u_i)M+D(u)M^{\frac13}\xi_i.
\label{xlscalingr23}
\end{align}
Here $A_2(u)$ is defined in~\eqref{A2}. Using~\eqref{zw0scale},
\eqref{xlscalingr23} and~\eqref{r3bq}, we get
\begin{align}
&\frac{z_1}{z_1-z_2}\sim
\frac{D(u)M^{\frac13}}{\t_1-\t_2-i(w_1+w_2)},\\
&\left(\frac{1-pz_1}{1-pz_2}\right)^n
\frac{1+1/z_2}{1+1/z_1}\sim 1,\\
&
\prod_{i=1}^{n}\frac{1-\bar{p}_iz_2}{1-\bar{p}_iz_1}
\sim
\begin{cases}
1,& \text{region}~2
\left(u\le u_c=\frac{\bar{q}^2-2q\bar{q}+q}{(\bar{q}-q)^2}\right),
\\
\prod_{i=1}^{n}\frac{\eta_i-\t_2-iw_2}{\eta_i-\t_1+iw_1},
&\text{region}~3~(u=u_c),
\end{cases}
\label{r2r3}
\\
&\frac{z_2^{x_2-\mu(u_2)M}}{z_1^{x_1-\mu(u_1)M}}
\sim z_c(u)^{D(u)M^{\frac13}(\xi_2-\xi_1)}
\exp(\xi_2\t_2-\xi_1\t_1+iw_1\xi_1+iw_2\xi_2).
\end{align}
Note that in~\eqref{r2r3}, the asymptotic form is different between
regions 2 and 3. This leads to the difference of the limiting 
distribution between Theorems 2-2 and 2-3'.

Thus we obtain the asymptotic form of 
$\tilde{K}(t_1, x_1;t_2, x_2)$, 
\begin{align}
&~\tilde{K}(t_1, x_1;t_2, x_2)\notag\\
&
\begin{cases}
\sim\frac{\nu}{D(u)M^{\frac13}}\int_0^{\infty}d\lambda e^{-\lambda(\t_1-\t_2)}
\Ai(\xi_1+\lambda)\Ai(\xi_2+\lambda), & \text{region}~2,\\
\sim\frac{\nu}{D(u)M^{\frac13}}\int_0^{\infty}d\lambda e^{-\lambda(\t_1-\t_2)}
\Ai(\xi_1+\lambda)\Ai(\xi_2+\lambda)\\
+\frac{\nu}{D(u)M^{\frac13}}
\sum_{j=1}^n\frac{1}{2\pi}\int_{-\infty}^{\infty}dw_1
\exp\left(i\xi_1w_1
+\frac{iw_1^3}{3}\right)\prod_{k=1}^j
\frac{1}{\eta_k-\t_1+iw_1}\\
\times
\frac{1}{2\pi}
\int_{-\infty}^{\infty}dw_2\exp\left(i\xi_2 w_2
+\frac{iw_2^3}{3}\right)\prod_{k=1}^{j-1}
(\eta_k-\t_2+iw_2), & \text{region}~3.
\end{cases}
\label{tKr23}
\end{align}
Here
\begin{equation}
\nu=\exp\left(M(f_{u_2}(z_c(u_2))-f_{u_1}(z_c(u_1)))+\xi_2\t_2
-\xi_1\t_1\right)
z_c(u)^{D(u)M^{\frac13}(\xi_2-\xi_1)}.
\label{c}
\end{equation}
In~\eqref{tKr23}, we used the integral representation of the Airy
function
\begin{equation}
\Ai(x)=\frac{1}{2\pi}\int_{-\infty}^{\infty}d\lambda e^{ix\lambda+\frac{i}{3}\lambda^3},
\label{Airyfunction}
\end{equation}
and the relation
\begin{align}
&~\frac{1}{\t_1-\t_2-i(w_1+w_2)}\left(\prod_{j=1}^N
\frac{\eta_j-\t_2-iw_2}{\eta_j-\t_1+iw_1}-1
\right)\notag\\
&=
\frac{1}{\eta_1-\t_1+iw_1}
+\frac{\eta_1-\t_2-iw_2}{(\eta_1-\t_1+iw_1)(\eta_2-\t_1+iw_1)}
+\cdots\notag\\
&\hspace{3cm}+
\frac{(\eta_1-\t_2-iw_2)(\eta_2-\t_2-iw_2)\cdots(\eta_{N-1}-\t_2-iw_2)}
{(\eta_1-\t_1+iw_1)(\eta_2-\t_1+iw_1)\cdots(\eta_N-\t_1+iw_1)}.
\label{stK}
\end{align}

Next we consider the asymptotics of 
$\phi_{t_1,t_2}(x_1,x_2)$~\eqref{thm1f}. Using $f_{u}(z)$~\eqref{fnr23} 
and $\mu(u)$~\eqref{mur23}, the function is rewritten as for 
$t_1<t_2$,
\begin{align}
\phi_{t_1,t_2}(x_1,x_2)=
\frac{1}{2\pi i}\int_{C_1} \frac{dz}{z} e^{M(f_{u_2}(z)-f_{u_1}(z))}
z^{x_2-\mu(u_2)M-x_1+\mu(u_1)M}.
\end{align}
Let the variable $z$ scale as
\begin{equation}
z=z_c(u)\left(1+\frac{i\sigma}{D(u)M^{\frac13}}\right),
\end{equation}
where $z_c(u)$ defined in~\eqref{dspr23} is the double saddle point 
of~\eqref{fnr23}. Due to~\eqref{r23crelation}, we have for $i=1,2$
\begin{equation}
z\sim z_c(u_i)\left(1+\frac{i\sigma-\t_i}{D(u)M^{\frac13}}
\right).
\end{equation}
From the above two equations,~\eqref{r23drelation} 
and~\eqref{xlscalingr23}, we get
\begin{align}
&e^{M(f_{u_2}(z)-f_{u_1}(z))}
\sim e^{M(f_{u_2}(z_c(u_2))-f_{u_1}(z_c(u_1)))}
e^{\frac{i}{3}(\sigma+i\t_2)^3-\frac{i}{3}(\sigma+i\t_1)^3}
\notag\\
&\hspace{2.8cm}
=e^{M(f_{u_2}(z_c(u_2))-f_{u_1}(z_c(u_1)))}e^{-(\t_2-\t_1)\sigma^2-
i(\t_2^2-\t_1^2)\sigma+\frac{\t_2^3-\t_1^3}{3}},\\
&z^{x_2-\mu(u_2)M-x_1+\mu(u_1)M}\sim
z_c(u)^{D(u)M^{\frac13}(\xi_2-\xi_1)}
e^{i\sigma(\xi_2-\xi_1)}.
\end{align}
Thus we obtain for $t_1<t_2$,
\begin{align}
&~\phi_{t_1,t_2}(x_1,x_2)\notag\\
&\sim\frac{\nu}{D(u)M^{\frac13}}
e^{\xi_1\t_1-\xi_2\t_2+\frac{\t_2^3-\t_1^3}{3}}
\int_{-\infty}^{\infty}d\sigma
e^{-(\t_2-\t_1)\sigma^2-
i(\t_2^2-\t_1^2)\sigma+i(\xi_2-\xi_1)\sigma}\notag\\
&=\frac{\nu}{D(u)M^{\frac13}}
\frac{1}{\sqrt{4\pi(\t_2-\t_1)}}
\exp\left(\frac{-(\xi_2-\xi_1)^2}{4(\t_2-\t_1)}
-\frac{(\xi_2+\xi_1)(\t_2-\t_1)}{2}+\frac{(\t_2-\t_1)^3}{12}\right),
\end{align}
where $\nu$ is given in~\eqref{c}. From~\eqref{Airyfunction} we finally 
find
\begin{equation}
\phi_{t_1,t_2}(x_1,x_2)
\sim\frac{\nu}{D(u)M^{\frac13}}
\int_{-\infty}^{\infty}d\lambda e^{-\lambda(\t_1-\t_2)}\Ai(\xi_1+\lambda)\Ai(\xi_2+\lambda).
\label{sphi}
\end{equation}
Hence from~\eqref{stK} and~\eqref{sphi} and noting that the term 
$\nu$~\eqref{c} does not affect the determinant, we finally obtain 
the desired kernels for the regions 2 and 3.

\subsection{Region 4 (proof of Theorem 2-4')}\label{pot24}
In this section, we discuss the asymptotics of the region 4 in Fig.~3. 
In this region we take time $t_j$ in such a way that they are 
macroscopically separated and the scaled time $u_j=t_j/M$ is taken as
$u_j> u_c$. The position $\ell_j$ of the tagged particle at $t_j$ and 
the stay rates $\bar{q}_i~(i=1,\cdots,n)$ of the defect particles are 
also scaled as~\eqref{12scaling} and~\eqref{r4bq} respectively.

Deforming the contour $C_{R_1}$ in the kernel $\tilde{K}(t_1,x_1;
t_2,x_2)$~\eqref{Ktxtx}, we divide it into two parts,
\begin{equation}
\tilde{K}(t_1,x_1;t_2,x_2)
=\tilde{K}_1(t_1,x_1;t_2,x_2)-\tilde{K}_2(t_1,x_1;t_2,x_2).
\end{equation}
Here for $i=1,2$
\begin{align}
&~\tilde{K}_i(t_1,x_1;t_2,x_2)\notag\\
&=-\frac{1}{(2\pi i)^2}\int_{\Gamma^{(i)}_{\bar{p}}}\frac{dz_1}{z_1}
\int_{C_{R_2}} \frac{dz_2}{z_2}\frac{z_1}{z_1-z_2}
\frac{(1+1/z_2)^{t_2-M+1}}{(1+1/z_1)^{t_1-M+1}}
\left(\frac{1-pz_2}{1-pz_1}\right)^{M-n}
\prod_{i=1}^n
\frac{1-\bar{p}_iz_2}{1-\bar{p}_iz_1}\frac{z_2^{x_2}}{z_1^{x_1}},
\label{r4K1}
\end{align}
where $p=q/(1-q)$ and $\bar{p}_i=\bar{q}_i/(1-\bar{q}_i)$. In the 
equation above, the contour $\Gamma^{(1)}_{\bar{p}}$ encloses 
$z=1/\bar{p}_i$ anticlockwise while the contour $\Gamma^{(2)}_{\bar{p}}$ 
is chosen in such a way that it encloses $z=-1,0,1/\bar{p}_i$ 
anticlockwise. 

First, we discuss the asymptotic form of $\tilde{K}_1(t_1,x_1;t_2,x_2)$. 
This is rewritten as
\begin{align}
\tilde{K}_1(t_1,x_1;t_2,x_2)
=-\frac{1}{(2\pi i)^2}\int_{\Gamma^{(1)}_{\bar{p}}}&\frac{dz_1}{z_1}
\int_{C_{R_2}}\frac{dz_2}{z_2}
e^{M\left(g_{u_2}(z_2)-g_{u_1}(z_1)\right)}\frac{z_1}{z_1-z_2}
\notag\\
&\times\left(\frac{1-pz_1}{1-pz_2}\right)^{n}
\prod_{i=1}^n\frac{1-\bar{p}_iz_2}{1-\bar{p}_iz_1}
\frac{1+1/z_2}{1+1/z_1}
\frac{z_2^{x_2-M\Lambda(u_2)}}
{z_1^{x_1-M\Lambda(u_1)}},
\label{r4K1'}
\end{align}
where $\Lambda(u)$ is a variable which will be fixed later and
\begin{equation}
g_u(z)=(u-1)\log(1+z)+\log(1-pz)+(\Lambda(u)-u+1)\log z.
\label{gnr23}
\end{equation}
We choose the value of $\Lambda(u)$ in such a way that the saddle point 
of $g_u(z)$ is $1/\bar{p}$ where $\bar{p}=\bar{q}/(1-\bar{q})$ . From 
the condition,
\begin{equation}
g'_{u}\left(\frac{1}{\bar{p}}\right)=0,
\label{r4sd}
\end{equation}
we have
\begin{equation}
\Lambda(u)=(u-1)-A_G(u)=\frac{\bar{p}(u-1)}{1+\bar{p}}
+\frac{p}{\bar{p}-p}.
\label{Lnr4}
\end{equation}
Scaling $z_i~(i=1,2)$ around the saddle point~\eqref{r4sd} as
\begin{equation}
z_i=\frac{1}{\bar{p}}\left(1-\frac{2w_i}
{D_G(u_i)M^{\frac12}}\right),
\label{zscalingr4}
\end{equation}
and noting $\bar{p}^2D_G^2(u)=2g''_{u}(1/\bar{p}),$ we can obtain the 
asymptotic form of $e^{M(g_{u_2}(z_2)-g_{u_1}(z_1))}$ in~\eqref{r4K1'} 
by the saddle point method,
\begin{align}
e^{M(g_{u_2}(z_2)-g_{u_1}(z_1))}\sim e^{M\left(g_{u_2}
\left(\frac{1}{\bar{p}}\right)
-g_{u_1}\left(\frac{1}{\bar{p}}\right)\right)}e^{z_2^2-z_1^2}.
\end{align}
Associated with~\eqref{12scaling}, we set
\begin{align}
x_i=(u_i-1)M-A_G(u_i)M+D_G(u_i)M^{\frac12}\xi_i,
\label{xscalingr4}
\end{align}
for $i=1,2$.
From this equation,~\eqref{zscalingr4} and~\eqref{r4bq}, we find
\begin{align}
&\prod_{i=1}^n\frac{1-\bar{p}_iw}{1-\bar{p}_iz}\sim
\prod_{i=1}^n\frac{e^{-\t_2}w_2+\e_i}{e^{-\t_1}w_1+\e_i},~~
\frac{z_1}{z_1-z_2}\sim \frac{M^{\frac12}}
{-2(e^{-\t_1}w_1-e^{-\t_2}w_2)},\notag\\
&\left(\frac{1-pz_1}{1-pz_2}\right)^{n}\frac{1+1/z_2}{1+1/z_1}
\sim 1,~~
\frac{z_2^{x_2-M\Lambda(u_2)}}{z_1^{x_1-M\Lambda(u_1)}}
\sim\left(\frac{1}{\bar{p}}\right)^{(e^{\t_2}\xi_2-e^{\t_1}\xi_1)
M^{\frac12}}e^{-2w_2\xi_2+2w_1\xi_1},
\end{align}
where we used the parameter $\t_i$ defined in~\eqref{r4tau}.
 
Thus we eventually get
\begin{align}
&~\tilde{K}_1(t_1,x_1;t_2,x_2)\notag\\
&\sim\frac{(1+\bar{p})^{t_2-t_1}
\bar{p}^{x_1-x_2}}
{e^{\t_2}M^{\frac12}}
\frac{2}{(2\pi i)^2}\int_{\Gamma}dw_1\int_{\g}dw_2
\frac{e^{w_2^2-w_1^2-2w_2\xi_2+2w_1\xi_1}}{e^{\t_1-\t_2}w_2-w_1}
\prod_{i=1}^n\frac{e^{-\t_2}w_2+\e_i}{e^{-\t_1}w_1+\e_i}.
\label{sK1r4}
\end{align}
Here the contour $\Gamma$ encloses $-e^{\t_1}\e_i~(i=1,\cdots,n)$
anticlockwise and $\g$ is an arbitrary path running from $-i\infty$ to 
$i\infty$.

Next we consider the asymptotics of 
$\phi_{t_1,t_2}(x_1,x_2)$~\eqref{thm1f}. By use of 
$g_{u}(z)$~\eqref{gnr23} and $\Lambda(u)$~\eqref{Lnr4}, we set for 
$t_1<t_2$,
\begin{equation}
\phi_{t_1,t_2}(x_1,x_2)
=\frac{1}{2\pi i}\int_{C_1} \frac{dz}{z} 
e^{M(g_{u_2}(z)-g_{u_1}(z))}
z^{x_2-x_1-\Lambda(u_2)+\Lambda(u_1)}.
\end{equation}
Scaling $z$ as
\begin{equation}
z=\frac{1}{\bar{p}}\left(1+i\frac{2w}{\sqrt{(D^2_G(u_2)-D^2_G(u_1))M}}
\right),
\end{equation}
From this equation and~\eqref{xscalingr4}, we get
\begin{align}
&e^{M(g_{u_2}(z)-g_{u_1}(z))}\sim e^{M\left(g_{u_2}
\left(\frac{1}{\bar{p}}\right)-
g_{u_1}\left(\frac{1}{\bar{p}}\right)\right)}
e^{-w^{2}},\\
&z^{x_2-x_1-M(\Lambda(u_2)-\Lambda(u_1))}\sim
\left(\frac{1}{\bar{p}}\right)^{(e^{\t_2}\xi_2-e^{\t_1}\xi_1)
M^{\frac12}}
\exp\left(2iw\frac{\xi_2-e^{\t_1-\t_2}\xi_1}
{\sqrt{1-e^{2(\t_1-\t_2)}}}\right),
\end{align}
Thus we obtain for $t_1<t_2$,
\begin{align}
\phi_{t_1,t_2}(x_1,x_2)&\sim\frac{\left(1+\bar{p}\right)^{t_2-t_1}
\bar{p}^{x_1-x_2}}{\pi\sqrt{(e^{2\t_2}-e^{2\t_1})M}}
\int_{-\infty}^{\infty}dw e^{-w^2+2iw\frac{\xi_2-e^{\t_1-\t_2}\xi_1}
{\sqrt{1-e^{2(\t_1-\t_2)}}}}\notag\\
&=\frac{\left(1+\bar{p}\right)^{t_2-t_1}\bar{p}^{x_1-x_2}}
{e^{\t_2}M^{\frac12}}\frac{e^{-\frac{(\xi_2-e^{\t_1-\t_2}\xi_1)^2}
{1-e^{2(\t_1-\t_2)}}}}{\sqrt{\pi(1-e^{2(\t_1-\t_2)})}}.
\label{sphir4}
\end{align}

At last, we discuss that $\tilde{K}_2(t_1,x_1;t_2,x_2)$ in \eqref{r4K1} 
does not contribute the asymptotic form of the kernel~\eqref{kernel}.
The derivation can be done in the similar fashion to the one in 
Theorem 3.1. in~\cite{IS2004}. Here we give only its outline. 

First we consider the case of equal time, $t_1=t_2=t=uM$. Scaling 
$z_i~(i=1,2)$ as~\eqref{zwscale}, we have
\begin{align}
&~\tilde{K}_2(t,x_1;t,x_2)\sim\frac{-z_c(u)^{D_G(u)M^{\frac12}
(\xi_2-\xi_1)}}{D(u)M^{\frac13}}\notag\\
&\times\int_{0}^{\infty}d\lambda \Ai\left(\frac{\Delta A(u)}{D(u)}M^{\frac23}
+\frac{D_G(u)}{D(u)}M^{\frac16}\xi_1+\lambda\right)
\Ai\left(\frac{\Delta A(u)}{D(u)}M^{\frac23}
+\frac{D_G(u)}{D(u)}M^{\frac16}\xi_2+\lambda\right),
\end{align}
where $\Delta A(u)=A_2(u)-A_G(u)$. Noticing that $\Delta A(u)>0$
for the region 4 ($u_c<u$) and the asymptotic form of the Airy 
function
\begin{equation}
\Ai(x)\sim \frac{1}{2\sqrt{\pi}}x^{-\frac14}
\exp\left(-\frac23x^{\frac32}\right),
\end{equation}
as $x\rightarrow\infty$, we find 
\begin{equation}
\tilde{K}_2(t,x_1;t,x_2)\sim e^{-\mathcal{O}(M)}.
\label{K2OTr4}
\end{equation}
This indicates that under the scaling~\eqref{12scaling}, the 
kernel $\tilde{K}_2(t,x_1;t,x_2)$ vanishes as $M$ goes to infinity.

Next we consider the case for arbitrary $t_i~(i=1,2)$. We scale $z_1$ 
around the double saddle point $z_c(u_1)$ of $f_{u_1}(z)$ and $z_2$ 
around the saddle point $1/\bar{p}$ of $g_{u_2}(z_1)$ as 
\begin{align}
z_1=z_c(u_1)\left(1-\frac{iw_1}{D(u_1)M^{\frac13}}\right),~
z_2=\frac{1}{\bar{p}}\left(1-\frac{2w_2}{D_G(u_2)M^{\frac12}}\right).
\end{align}
Under these scalings, we find
\begin{align}
\tilde{K}_2(t,x_1;t,x_2)\sim -\frac{(1+\bar{p})^{t_2-t_1}
\bar{p}^{x_1-x_2} e^{-\xi_2^2}D_G(u_1)}{\sqrt{\pi}D_G(u_2)}\times
\Theta(u_1,\xi_1),
\label{r4K2}
\end{align}
where
\begin{align}
&\Theta(u_1,\xi_1)\notag\\
=&\frac{1}{z_c(u_1)-1/\bar{p}}\left(\frac{1-p/\bar{p}}{1-pz_c(u_1)}
\right)^{M-n}\prod_{i=1}^n\frac{1-\bar{p}_i/\bar{p}}{1-\bar{p}_iz_c(u_1)}
\frac{z_c(u_1)^{-x_1}\bar{p}^{-x_1}}{D(u_1)D_G(u_1)M^{\frac56}\bar{p}}
\notag\\
&\times\frac{(1+\bar{p})^{(u_1-1)M+1}}{\left(1+\frac{1}{z_c(u_1)}
\right)^{(u_1-1)M+1}}
\Ai\left(\frac{\Delta A(u_1)}{D(u_1)}M^{\frac23}
+\frac{D_G(u_1)}{D(u_1)}M^{\frac16}\xi_1\right).
\end{align}
Considering the case $t_1=t_2$ in~\eqref{r4K2} and the former 
result~\eqref{K2OTr4}, one easily finds
\begin{equation}
\Theta(u,\xi)\sim e^{-\mathcal{O}(M)}.
\end{equation}
Hence we finally get
\begin{equation}
\tilde{K}_2(t_1,x_1;t_2,x_2)
\sim
(1+\bar{p})^{t_2-t_1}\bar{p}^{x_1-x_2}
\left(e^{-\mathcal{O}(M)}\right).
\label{sK2r4}
\end{equation}

From~\eqref{sK1r4},~\eqref{sphir4} and~\eqref{sK2r4} and noting the 
factor $(1+\bar{p})^{t_2-t_1}\bar{p}^{x_1-x_2}$ does not affect the 
determinant, we obtain the desired expression for the limiting kernel.

\subsection{Fixed $M$ case (proof of Theorem 3)}\label{pot3}
In this section we discuss the scaling limit explained in 
Section~\ref{sclim2} where time $t$ goes to infinity with $M$ fixed.
In this limit, we scale the time $t_j$, the particle position $\ell_j$,
and the stay rates $q_i~(i=1,\cdots, M)$ 
as~\eqref{r5t},~\eqref{r5elli}, and~\eqref{r5qi} respectively.

First we consider the scaling form of 
$\tilde{K}(t_1,x_1;t_2,x_2)$~\eqref{Ktxtx}.
Changing the variables $z_i~(i=1,2)$ to $z'_i=1/z_i$, we have
\begin{equation}
\tilde{K}(t_1,x_1;t_2,x_2)
=\frac{1}{(2\pi i)^2}\int_{\Gamma_p}\frac{dz'_1}{z'_1}\int_{C_{R_1}} 
\frac{dz'_2}{z'_2}\frac{z'_2}{z'_2-z'_1}
\frac{(1+z'_2)^{t_2-M+1}}{(1+z'_1)^{t_1-M+1}}\prod_{i=1}^M
\frac{z'_2-p_i}{z'_1-p_i}\frac{{z'_1}^{x_1+M}}
{{z'_2}^{x_2+M}},
\end{equation}
where the contour $\Gamma_p$ encloses $p_i=q_i/(1-q_i)$ anticlockwise, 
and $C_{R_1}$ encloses the origin $z'$ and $p_i=q_i/(1-q_i)$. 
Introducing $y(\t_i)~(i=1,2)$, which will be chosen later, we set 
\begin{align}
&~\tilde{K}(t_1,x_1;t_2,x_2)\notag\\
&=\frac{1}{(2\pi i)^2}\int_{\Gamma_p}\frac{dz'_1}{z'_1}\int_{C_{R_1}} 
\frac{dz'_2}{z'_2}
\frac{z'_2}{z'_2-z'_1}e^{T\left(h_{\t_2}(z'_2)-h_{\t_1}(z'_1)\right)}
\prod_{i=1}^M
\frac{z'_2-p_i}{z'_1-p_i}\frac{{z'_1}^{x_1-y(\t_1)T+M}}
{{z'_2}^{x_2-y(t_2)T+M}}\left(\frac{1+z'_1}{1+z'_2}\right)^{M-1},
\end{align}
where the parameter $\t$ is defined in~\eqref{r5t} and
\begin{equation}
h_{\t}(z)=e^{2\t}\ln(1+z)-y(\t)\ln(z).
\label{r5h}
\end{equation}
We choose $y(\t)$ in a way that $h_{\t}(z)$ has the saddle point at 
$z=p=q/(1-q)$ where $q$ is defined in~\eqref{r5qi}. Thus from the 
condition, $h'_{\t}(p)=0$, we have
\begin{equation}
y(\t)=\frac{pe^{2\t}}{1+p}.
\label{r5y}
\end{equation}
We scale the variable $z'_i~(i=1,2)$ around the saddle point $z_c=p$ 
as
\begin{align}
z'_i=p\left(1+\frac{\sqrt{2}(1+p)}{e^{\t_i}{(pT)}^{\frac12}}w_i
\right).
\end{align}
Considering~\eqref{r5elli}, we also scale $x_i~(i=1,2)$ as
\begin{align}
x_i=\frac{pe^{2\t_i}}{1+p}T+\xi_i\frac{e^{\t_i}\sqrt{2pT}}{1+p}.
\label{xscaling3}
\end{align}
From these equations and~\eqref{r5qi}, we find 
\begin{align}
&e^{T\left(h_{\t_2}(z'_2)-h_{\t_1}(z'_1)\right)}
\sim e^{T\left(h_{\t_2}(p)-h_{\t_1}(p)\right)}e^{w_2^2-w_1^2},~~
\frac{z'_2}{z'_2-z'_1}\sim \frac{1}{e^{-\t_2}w_2-e^{-\t_1}w_1}
\frac{(pT)^{\frac12}}{\sqrt{2}(1+p)},\notag\\
&\frac{{z'_1}^{x_1-y(\t_1)T}}{{z'_2}^{x_2-y(\t_2)T}}
\sim p^{\frac{\sqrt{2pT}}{1+p}(e^{\t_1}\xi_1-e^{\t_2}\xi_2)}
e^{2w_1\xi_1-2w_2\xi_2},~~
\prod_{i=1}^M\frac{z'_2-p_i}{z'_1-p_i}\sim\prod_{j=1}^M
\frac{e^{-\t_2}w_2+\e_j}{e^{-\t_1}w_1+\e_j},\notag\\
&\frac{{z'_1}^{M}}{{z'_2}^M}\left(\frac{1+z'_1}{1+z'_2}\right)^{M-1}
\sim 1.
\end{align}
From these equations, we eventually obtain
\begin{align}
&~\tilde{K}(t_1,x_1;t_2,x_2)\notag\\
&\sim
(1+p)^{t_2-t_1}p^{x_1-x_2}\frac{(1+p)}{e^{\t_2}\sqrt{2pT}}
\frac{2}{(2\pi i)^2}\int_{\Gamma}dw_1\int_{\g}dw_2
\frac{e^{w_2^2-2w_2\xi_2-w_1^2+2w_1\xi_1}}{e^{\t_1-\t_2}w_2-w_1}
\prod_{j=1}^M\frac{e^{-\t_2}w_2+\e_j}{e^{-\t_1}w_1+\e_j}.
\label{tK3}
\end{align}
Here the contour $\Gamma$ encloses $-e^{\t_1}\e_i,~(i=1,\cdots,M)$
anticlockwise and $\g$ is arbitrary path from $-i\infty$ to $i\infty$. 

Next we consider the scaling limit of 
$\phi_{t_1,t_2}(x_1,x_2)$~\eqref{thm1f}.
Changing the variable $z$ to $z'=1/z$, we set
\begin{equation}
\phi_{t_1,t_2}(x_1,x_2)
=\frac{1}{2\pi i}\oint \frac{dz'}{z'} e^{T(h_{\t_2}(z')-h_{\t_1}(z'))}
z^{x_1-x_2-\left(y{(\t_1)}-y{(\t_2)}\right)T},
\end{equation}
where $h_{\t}(z)$ and $y(\t)$ are given in~\eqref{r5h} and
\eqref{r5y}. We set
\begin{equation}
z'=p\left(1+\sqrt{\frac{2}{(e^{2\t_2}-e^{2\t_1})pT}}(1+p)w\right).
\end{equation}
From this equation and~\eqref{xscaling3}
\begin{align}
&e^{T(h_{\t_2}(z)-h_{\t_1}(z))}
\sim e^{T(h_{\t_2}(z)-h_{\t_1}(z))}e^{w^2},\notag\\
&z^{x_1-x_2-(y_{\t_1}-y_{\t_2})T}
\sim
p^{(\t_2\xi_2-\t_1\xi_1)\frac{\sqrt{2pT}}{1+p}}
\exp\left(\frac{2(\xi_2-\xi_1)w}{\sqrt{e^{2\t_2}-e^{2\t_1}}}
\right).
\end{align}
Combining these equations, we finally find
\begin{align}
&~\phi_{t_1,t_2}(x_1,x_2) \notag\\
&\sim
(1+p)^{t_2-t_1}p^{x_1-x_2}
\frac{\sqrt{2}(1+p)}{\sqrt{(e^{2\t_2}-e^{2\t_1})Tp}}
\frac{1}{2\pi i}\int_{-\infty}^{\infty} dw 
\exp\left(w^2+\frac{2(\xi_2-\xi_1)w}
{\sqrt{e^{2\t_2}-e^{2\t_1}}}
\right),\notag\\
&=(1+p)^{t_2-t_1}p^{x_1-x_2}
\frac{1+p}{e^{\t_2}\sqrt{2pT}}
\sqrt{\frac{1}{\pi(1-e^{2(\t_2-\t_1)})}}
\exp\left(\frac{-(\xi_2-e^{\t_1-\t_2}\xi_1)^2}
{1-e^{2(\t_1-\t_2)}}\right).
\label{phi3}
\end{align}
Thus from~\eqref{tK3} and~\eqref{phi3} and noting that the prefactor
$(1+p)^{t_2-t_1}p^{x_1-x_2}$ does not affect the determinant, we get 
the desired expression for the kernel.

\section{Discussion}\label{discussion}
\subsection{Numerical simulations}
In this section, we give the result of Monte-Carlo simulations about 
the position fluctuation of a tagged particle in the TASEP in order to 
check our analysis discussed in the preceding sections.

We performed the simulations in two situations, the case without 
defect particle and that with one defect particle. Fig.~8 shows the 
data of the scaled position of a tagged particle  obtained by 
the Monte-Carlo simulation and the probability distribution functions 
which must fit them.  Fig.~8(a) corresponds to the time region 
$1/(1-q)<u<u_c=(\bar{q}^2-2q\bar{q}+q)/(\bar{q}-q)^2$ where $q$ (resp.
$\bar{q}$) is the stay rate of the normal particle (resp. the defect
particle). Note that in the region, the both situations belong to the
region 2 and due to Theorem 2-2, the GUE Tracy-Widom distribution 
must fit the data. Fig.~8(b) represents the case $u=u_c$. In this case
the first situation comes under the region 2. On the other hand, 
the second one belongs to the region 3 where the position fluctuation 
are described by the limiting largest eigenvalue distribution of GOE$^2$ 
as explained in Theorem 2-3. In Fig.~8(c), only the data for the second 
situation are shown for the case $u_c < u$.  This case is classified as 
the region 4 and from Theorem 2-4, the fluctuation is supposed to be 
Gaussian. In all figures, we see a good agreement between the data and 
the distribution functions which they must obey.    

\subsection{Correlation of current fluctuations}
In this article, we have discussed the multi-time distribution of the 
tagged particle~\eqref{1}. In order to analyze it, we have introduced 
the directed polymer problem of the 01 matrix in Section~\ref{ftts}. 

Here we consider the directed polymer problem of another random matrix 
where each element is geometric random variable. By this analysis 
we can discuss the correlations of other quantities in the TASEP with 
the step initial condition.

Let $\{a_g(i,j)\}_{1\le i\le N, 1\le j\le M}$ be the $N\times M$
matrix and the element $a_g(i,j)$ is the geometric random variable,
\begin{equation}
\text{Prob}(a_g(i,j)=k)=(1-q_j)q_j^{k},
\end{equation}
where $q_j$ is the parameter of the geometric distribution and we 
identified this with the stay rate of the $j$th particle of the TASEP
with the step initial condition defined in Section~\ref{model}.
For the matrix, we introduce the quantity $G^*_g(N,M)$ analogous to 
$G(N,M)$~\eqref{defpi} as
\begin{equation}
G^*_g(N,M)=N+M-1+\max_{\pi_g(N,M)}\sum_{(i,j)\in\pi_g(N,M)}a_g(i,j).
\end{equation}
Here $\pi_g(N,M)$ is the set of right/down paths from $(1,1)$ to 
$(N,M)$,
\begin{align}
\pi_g(N,M)&=\left\{\{(i_k,j_k)_{k=1,2,\cdots,N+M-1}\}|
(i_1,j_1)=(1,1),(i_{N+M-1},j_{N+M-1})=(N,M),\right.\notag\\
&~~~\left.(i_{k+1}-i_{k},j_{k+1}-j_{k})=(1,0) \text{~or~}
(0,1)
\right\}.
\end{align}
In this setting, we consider the following quantity,
\begin{equation}
\text{Prob}\left(G^*_g(N_1,M)\le t_1,\cdots, G^*_g(N_m,M)\le t_m\right).
\label{probG}
\end{equation}

It is known that $G^*_g(N,M)$ can be interpreted as a quantity in 
the TASEP as follows~\cite{Jo2000,Ba2006},
\begin{quote}
$G^*_g(N,M)=t$: In the TASEP with the step initial condition, the 
time until which the $M$th particle moves $N$ sites to its right is $t$.
\end{quote} 
Thus the quantity~\eqref{probG} represents the correlations of the 
arrival times of the $M$th particle at the site $N_i~(i=1,\cdots,m)$. 
(Note that we set the site coordinate as in Fig.~1(b).) Furthermore we 
easily find that
\begin{equation}
\text{Prob}\left(G^*_g(N,M)\le t\right)
=
\begin{cases}
\text{Prob}\left(H(t,N)\ge N\right), & \text{for~} N\le M,\\
\text{Prob}\left(H(t,N)= M\right), & \text{for~} M\le N. 
\end{cases}
\end{equation}
Here $H(t,N)$ represents the number of particles which passed the site 
$N$ until time $t$. Thus the probability~\eqref{probG} also means the 
correlations of currents between different times and different sites.

Applying the similar technique in Section~\ref{ftts} to~\eqref{probG}, 
we can also represent it as the growth process of Young diagram 
characterized by the Schur process. The result is as follows,
\begin{align}
&~\text{Prob}\left(G^*_g(N_1,M)\le t_1,\cdots, G^*_g(N_m,M)\le t_m
\right)
\notag\\
&=\text{Prob}\left(\lambda_1(N_1,M)\le t_1-N-M+1,\cdots, 
\lambda_1(N_m,M)\le t_m-N-M+1\right),
\label{GtSchur}
\end{align}
where $\lambda(N_i,M)$ is the Young diagram obtained by applying the normal 
RSK algorithm to the submatrix $\{a_g(i,j)\}_{i=1,\cdots,N_i, j=1,
\cdots,M}$ and $\lambda_1(N_i,M)$ is the length of its first row. The 
probability measure in the right hand side of this equation is 
characterized by the following joint distribution function,
\begin{align}
&\text{Prob}\left(\lambda(1,M)=\lambda^{(1)},\cdots, \lambda(N,M)=\lambda^{(N)}\right)
\notag\\
&=s_{\lambda^{(N)}}(q_1,q_2,\cdots,q_M)s_{\lambda^{(1)}}(1,0,\cdots)
s_{\lambda^{(2)}/\lambda^{(1)}}(1,0,\cdots)\cdots 
s_{\lambda^{(N)}/\lambda^{(N-1)}}(1,0,\cdots)\prod_{i=1}^M(1-q_i)^N,
\end{align} 
where $s_{\lambda/\mu}(q_1,q_2,\cdots)$ is the Schur function. This is also 
a special case of the Schur process and we can get the Fredholm 
determinant representation of~\eqref{GtSchur}.  Although in one point 
case, the asymptotics of this equation was discussed 
in~\cite{Jo2000,Ba2006}, the multi-point distribution has not been 
discussed yet.  

\section{Conclusion}\label{conclusion}
In this article, we have studied the multi-time distribution
function~\eqref{1} of position fluctuations of a tagged particle 
in the TASEP with the step initial condition. The main results are
summarized as Theorems 1, 2 and 3 in Section~\ref{main}. 

First, we have obtained the Fredholm determinant expression of~\eqref{1} 
in Theorem 1. For this purpose, we have mapped the time evolution of a 
tagged particle in the TASEP to the growth process of Young diagram 
which is related to the special case of the Schur process. Next, using 
the Fredholm determinant in Theorem 1, we have studied the two types of 
scaling limit. The first one is the case where both time $t$ and the 
label of a tagged particle $M$ go to infinity. In the second one, we 
take the $t\to\infty$ limit with $M$ fixed. The results for the first and 
second ones are shown in Theorems 2 and 3 respectively in 
Section~\ref{main}. 

In the first scaling limit, if the hopping rates of all particles are 
the same , we can divide the scaled time into two characteristic regions 
according to the limiting behavior of~\eqref{1}, the region 1 where a 
tagged particle begins to move and the region 2 which is after the 
region 1. In the region 1, the limiting process of a tagged particle 
converges to the spatially discrete process which is described in 
Theorem 2-1. The process reflects on the discreteness of the model. 
In the region 2, on the other hand, we have shown in Theorem 2-2 that 
the limiting process becomes the Airy process, which is characteristic 
of the one-dimensional KPZ universality class. 

If there are $n$ defect particles with small hoping rates in front of a 
tagged particle, the limiting distribution changes at the scaled time
$u_c$ which is determined by the hopping rate of the slowest defect 
particle. The limiting process around $u_c$ (region 3) is equivalent to 
that of the largest eigenvalue in GUE Dyson's Brownian motion with rank 
one external source as described in Theorem 2-3. Theorem 2-3' indicates
that the rank becomes $n$ if the hopping rates of the defect particles
are the same. When the scaled time is after $u_c$ (region 4), we have 
found in Theorem 2-4 that the process is equivalent to the one 
dimensional Brownian motion. If the hopping rates are degenerate, it is 
equivalent to the process of the largest eigenvalue of $n\times n$ GUE 
Dyson's Brownian motion model as described in Theorem 2-4'. This 
indicates that in the region 4, the effect of the defect particles is 
dominant whereas that of an infinite number of normal particles is 
irrelevant.

Theorem 3 shows the result for the second scaling limit. The limiting 
distribution is also described as the Fredholm determinant whose kernel 
is the same as that in Theorem 2-4'.

\section*{Acknowledgments}
The authors would like to thank A. Borodin for drawing out attention
to the reference~\cite{BO2006c}. The work of T.I. is supported by Core 
Research for Evolutional Science and Technology of Japan Science and 
Technology Agency. The work of T.S. is supported by the Grant-in-Aid 
for Young Scientists (B), the Ministry of Education, Culture, Sports, 
Science and Technology, Japan.

\newpage
\begin{large}
\noindent
Figure Captions
\end{large}

\vspace{10mm}
\noindent
Fig.~1: 
Definition of the TASEP with the step initial condition. (a) During 
each time step, the $i$th particle can hop to the right with probability 
$1-q_i$. However, $i+2$th particle cannot hop to the right neighboring 
site since the site is occupied by $i+1$th particle. (b) Step 
initial condition.

\vspace{10mm}
\noindent
Fig.~2:
Typical time evolution of the particles from the first to 100th in 
the TASEP with the step initial condition. The hopping rate of all 
particles is 0.9 ($q=0.1$) except the four defect particles (the first, 
25th, 50th and 75th particles) whose hopping rate is 0.8 
($\bar{q}=0.2$). Fig.~(a) illustrates  the whole time evolution from $t=0$ 
through 3000. Its closeup around $t=200$ and $t=3000$ is shown in (b) 
and (c) respectively. 
 
\vspace{10mm}
\noindent
Fig.~3:
Average position of the tagged particle. The thick line 
shows~\eqref{average}. We divide this line into four regions, where $u$ 
is near $1/(1-q)$, $1/(1-q)<u<u_c$, $u$ is near $u_c$, and $u_c<u$. 
We denote them as region 1, 2, 3, and 4 respectively.

\vspace{10mm}
\noindent
Fig.~4:
Example of the time evolution of the TASEP and corresponding 01 matrices. 
In this case, two tables is assigned for one example of the time 
evolution. Since the second particle can not hop to its right 
neighboring site between $t=2$ and $3$, there are two possibilities of 
assigning both 0 and 1 in the $(i,j)=(1,2)$ element of the 01 matrix. 

\vspace{10mm}
\noindent
Fig.~5:
(a) Example of time evolution of the TASEP and (b) one of its 
corresponding matrices. Note that a set of many matrices corresponds 
to (a) and only one of them is shown here.

\vspace{10mm}
\noindent
Fig.~6:
(a) Generalized permutation corresponding to Fig.~5(b). We arrange 
$(i,j)$ where $a(i,j)=1$ following the rule in~\cite{St1999}.
(b) The pair $(P,Q)$ obtained from (a) by the dual RSK algorithm.

\vspace{10mm}
\noindent
Fig.~7:
Growth of Young diagram obtained from Fig.~5(b) by the dual RSK 
algorithm. The diagrams $\lambda(i,4)~(1=1,\cdots,6)$ are obtained from the 
submatrices $\{a(j,k)\}_{j=1,\cdots,i,k=1,\cdots,4}$ of Fig.~5(b).
The set of the diagrams $\{\lambda(i,4)\}_{i=1,\cdots,6}$ corresponds to
the $Q$ tableaux in Fig.~6(b).

\vspace{10mm}
\noindent
Fig.~8:
Probability distributions of the scaled position of the 100th particle 
from the right ($M=100$) for $t=200$ (a), $t=1000$ (b) and $t=3000$ (c). 
In these figures, $\times$ represents the data for the case without 
defect particles and we set the stay rate $q=0.1$. On the other hand, 
$+$ corresponds to the case where the first particle is a defect 
particle with $\bar{q}=0.2$ while remaining particles are normal ones 
with $q=0.1$. The number of samples are 10000 for each case.
In (a), both cases belong to the region 2 and are fitted 
into the GUE Tracy-Widom distribution shown as the dashed line. In (b), 
the second case belongs to the region 3 since $u=t/M=10=u_c$ where 
$u_c=(\bar{q}^2-2\bar{q}q+q)/(\bar{q}-q)^2$ and it is described by 
the distribution denoted as GOE$^2$ while the first case remains in
region 2. In (c), the second cases comes under the region 4 where
the distribution is described by Gaussian (dashed line).    

\newpage

\noindent 
Fig.~1

\begin{picture}(400,171)
\put(0,160){(a)}
\psfrag{1-q_i}{$1-q_i$}
\psfrag{1-q_i+1}{$1-q_{i+1}$}
\put(0,100){\includegraphics{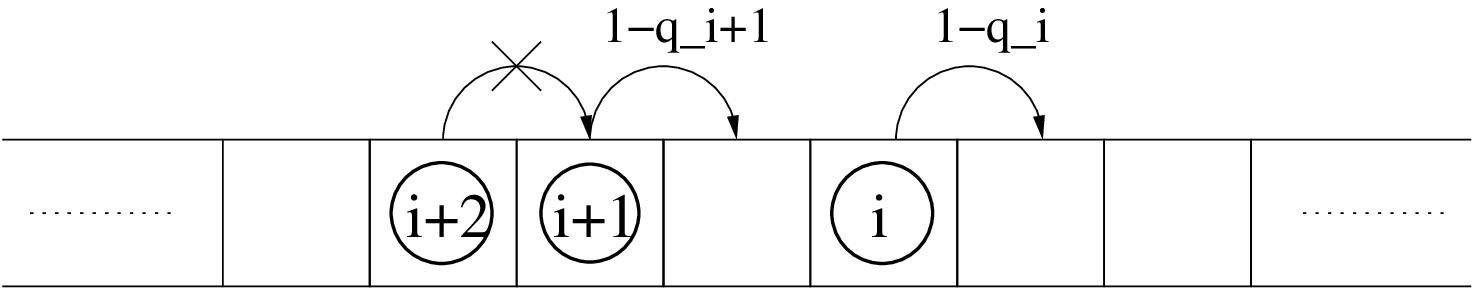}}
\put(0,75){(b)}
\put(0,0){\includegraphics{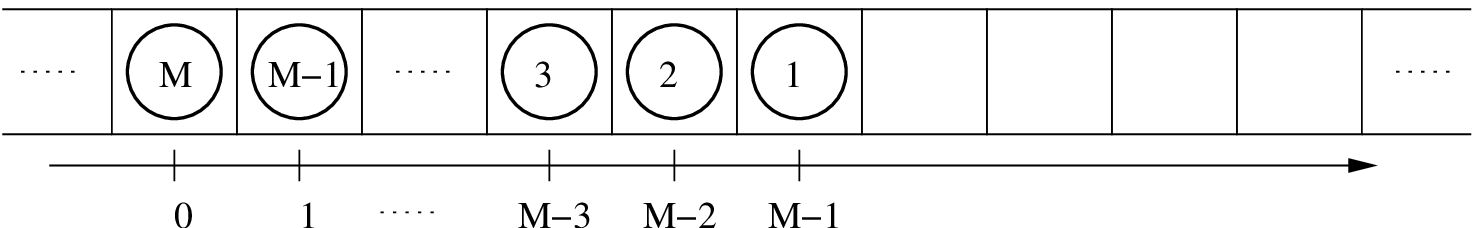}}
\end{picture}

\vspace{10mm}

\noindent
Fig.~2

\begin{picture}(400,171)
\put(0,160){(a)}
\put(220,160){(b)}
\put(110,0){time}
\put(330,0){time}
\put(15,155){position}
\put(240,155){position}
\put(0,5){\scalebox{0.6}{\includegraphics{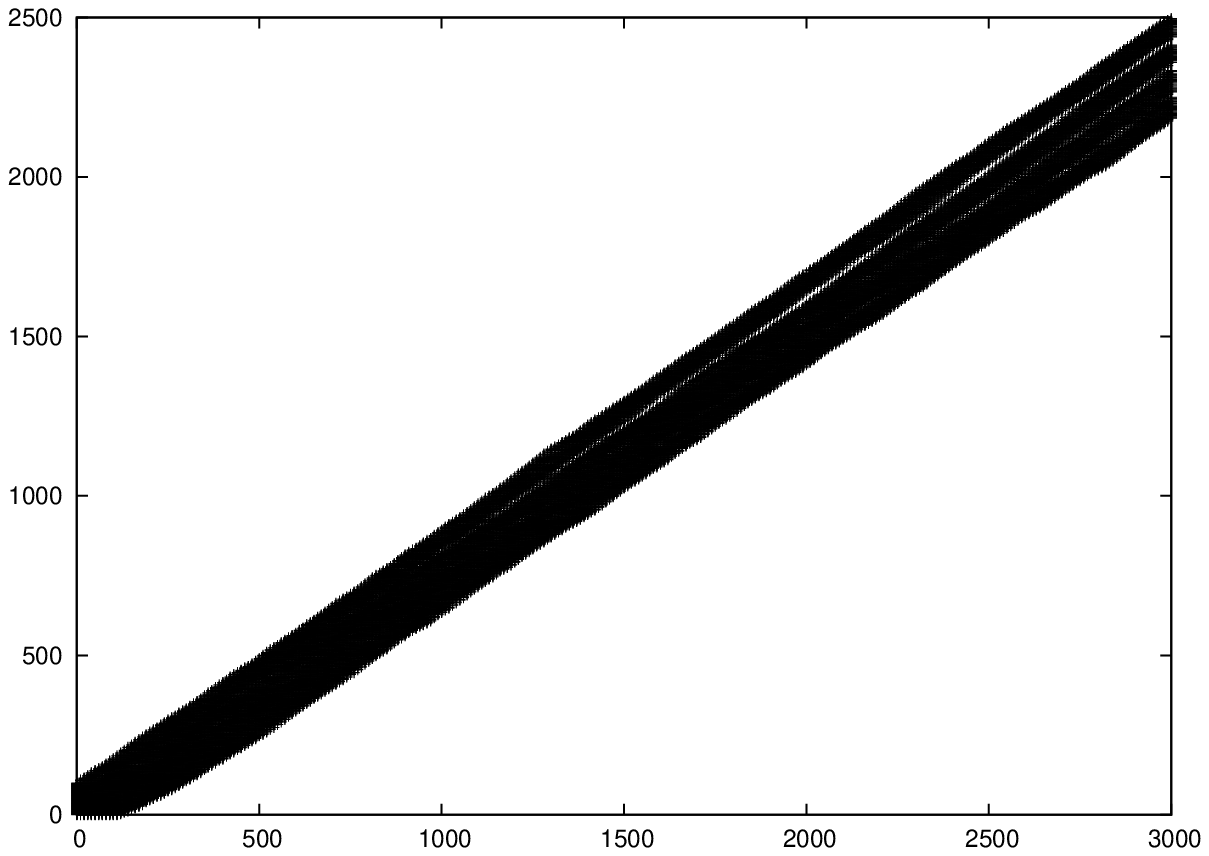}}}
\put(220,5){\scalebox{0.6}{\includegraphics{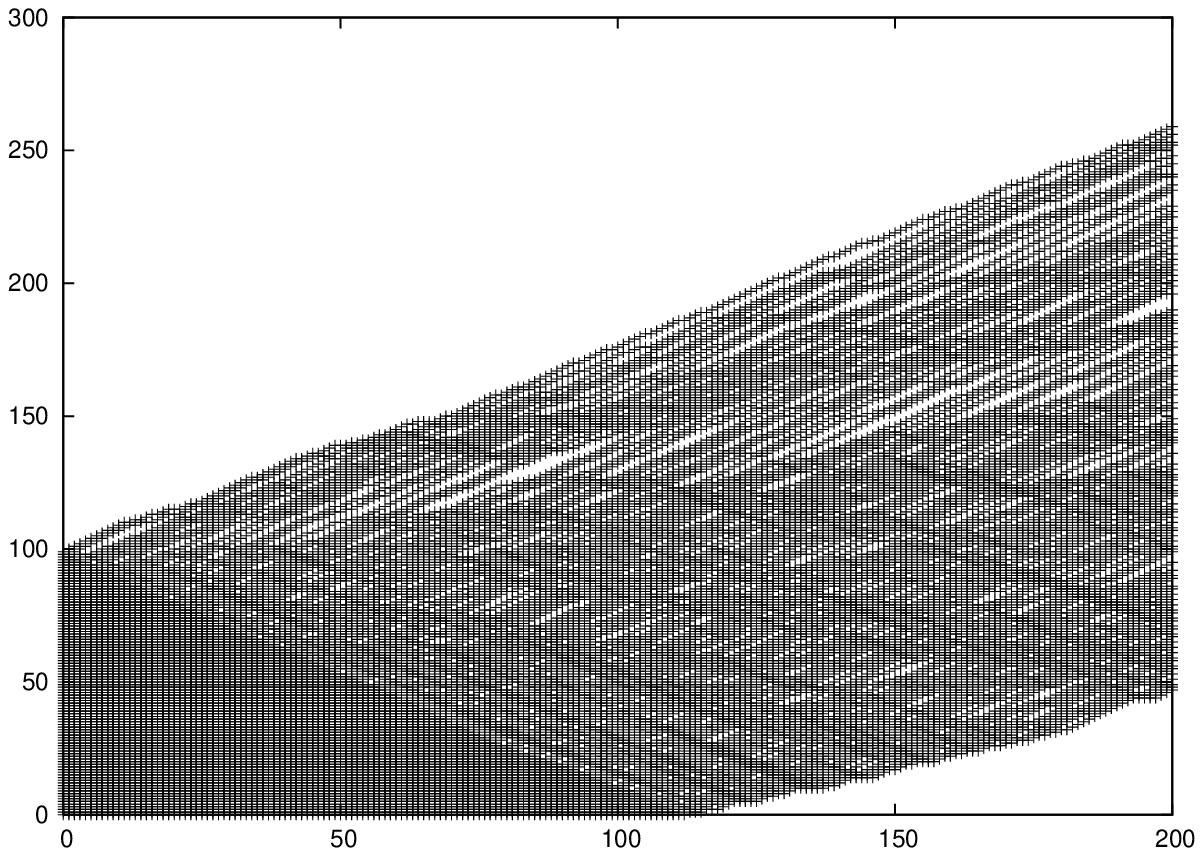}}}
\end{picture}

\begin{picture}(400,171)
\put(0,160){(c)}
\put(110,0){time}
\put(15,155){position}
\put(0,5){\scalebox{0.6}{\includegraphics{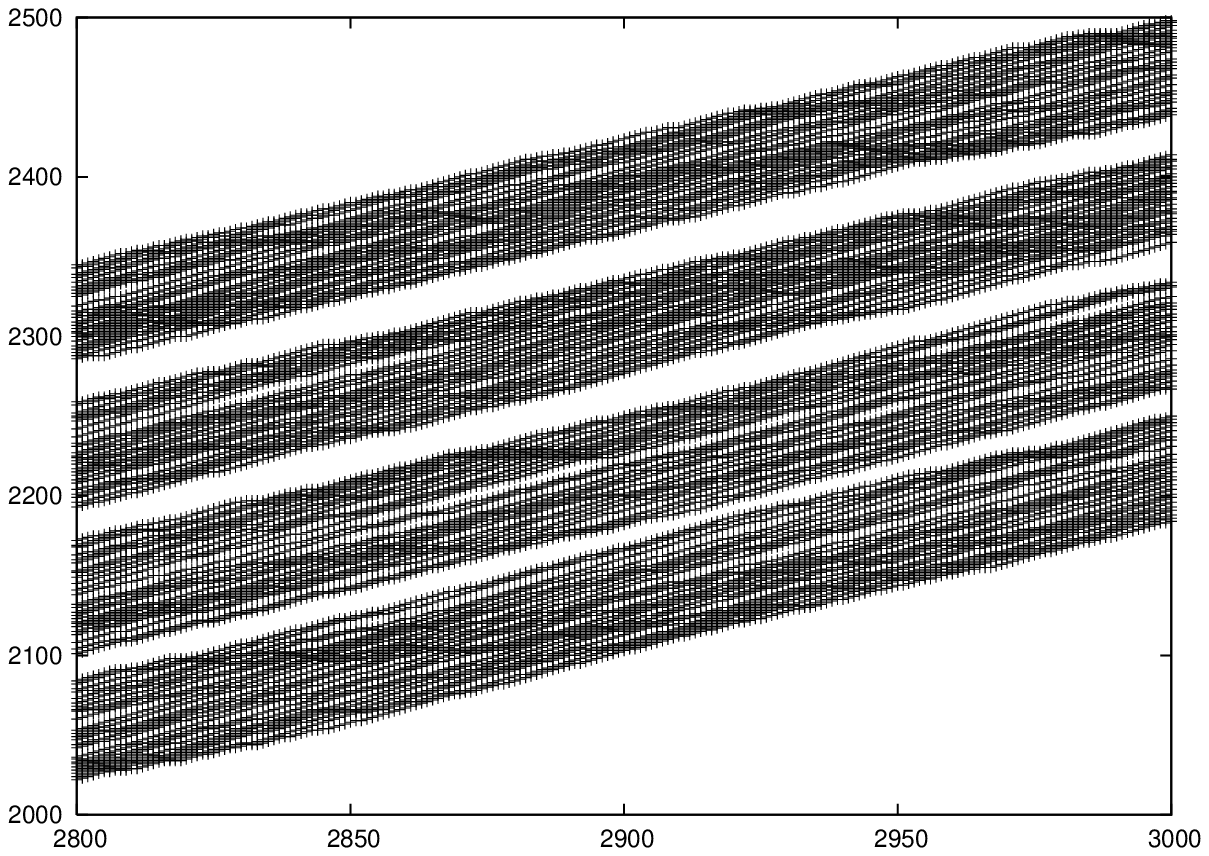}}}
\end{picture}

\newpage

\noindent
Fig.~3

\begin{picture}(400,171)
\label{tspand01}
\psfrag{A}{$A(u)$}
\psfrag{ts}{$u$}
\psfrag{q}{$\frac{1}{1-q}$}
\psfrag{tc}{$u_c$}
\put(0,0){\scalebox{0.8}{\includegraphics{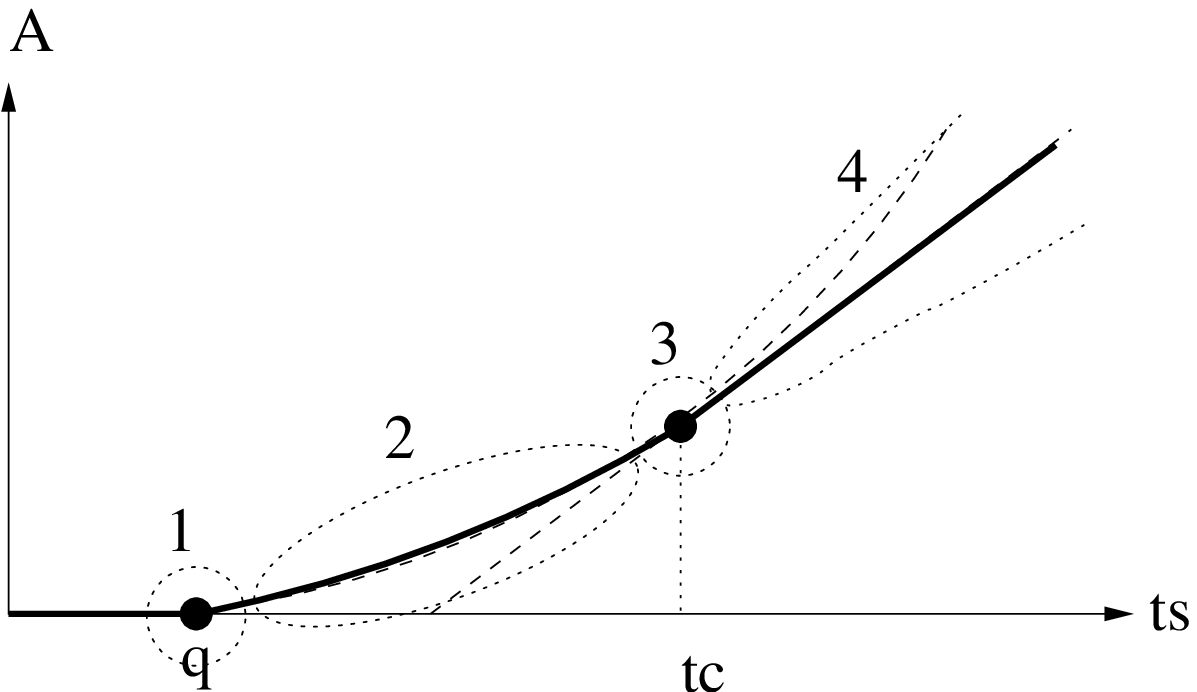}}}
\end{picture}

\vspace{10mm}

\noindent
Fig.~4

\begin{picture}(400,171)
\put(0,130){$t=0$}
\put(105,130){$t=1$}
\put(210,130){$t=2$}
\put(320,130){$t=3$}
\put(95,30){$i=1$}
\put(95,10){$i=2$}
\put(110,50){$j=1$}
\put(150,50){$2$}
\put(0,0){\scalebox{0.6}{\includegraphics{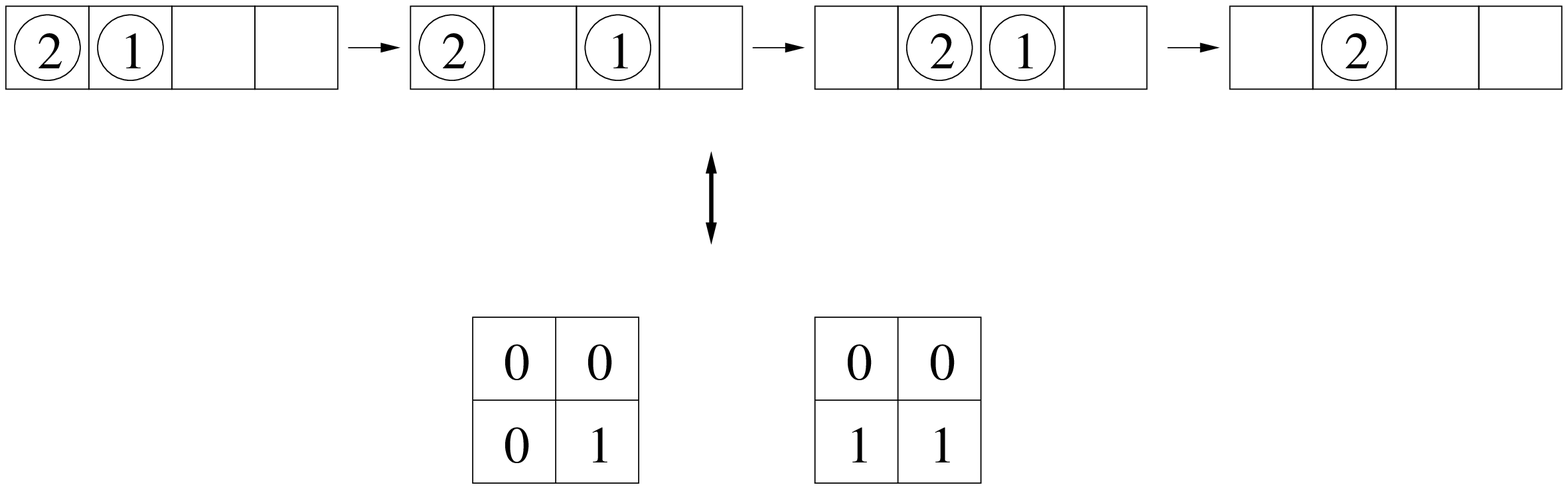}}}
\end{picture}

\noindent 
Fig.~5

\begin{picture}(400,171)
\put(0,160){(a)}
\put(65,171){$t=0$}
\put(160,171){$t=1$}
\put(257,171){$t=2$}
\put(65,147){$t=3$}
\put(160,147){$t=4$}
\put(257,147){$t=5$}
\put(65,123){$t=6$}
\put(160,123){$t=7$}
\put(257,123){$t=8$}
\put(65,98){$t=9$}
\put(50,85){\scalebox{0.4}{\includegraphics{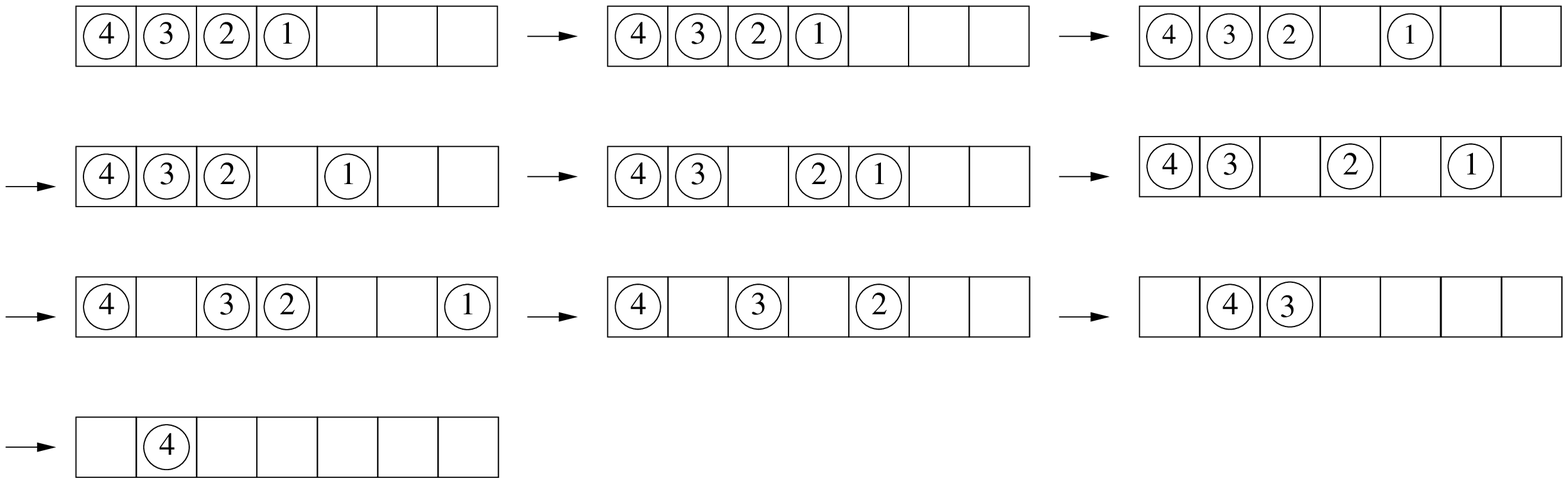}}}
\put(0,70){(b)}
\put(70,55){$i=1$}
\put(70,40){$i=2$}
\put(70,26){$i=3$}
\put(70,11){$i=4$}
\put(70,-3){$i=5$}
\put(70,-18){$i=6$}
\put(82,70){$j=1$}
\put(118,70){$2$}
\put(133,70){$3$}
\put(147,70){$4$}
\put(100,-20){\scalebox{0.5}{\includegraphics{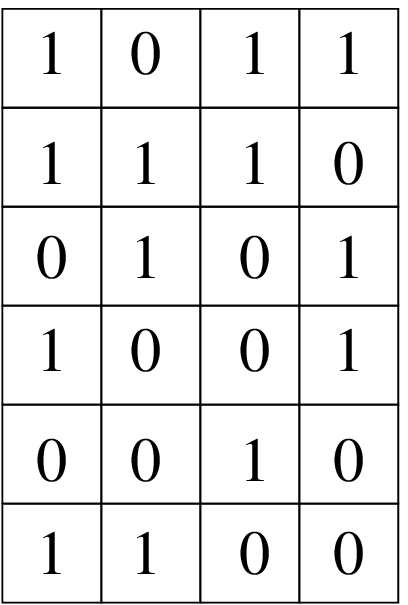}}}
\end{picture}

\vspace{10mm}

\noindent
Fig.~6

\begin{picture}(400,171)
\put(0,160){(a)}
\put(5,130){$i$}
\put(5,120){$j$}
\put(15,125){$\Big($}
\put(20,130){1~1~1~2~2~2~3~3~4~4~5~6~6}
\put(20,120){1~3~4~1~2~3~2~4~1~4~3~1~2}
\put(143,125){$\Big)$}
\put(200,160){(b)}
\put(200,40){\scalebox{0.5}{\includegraphics{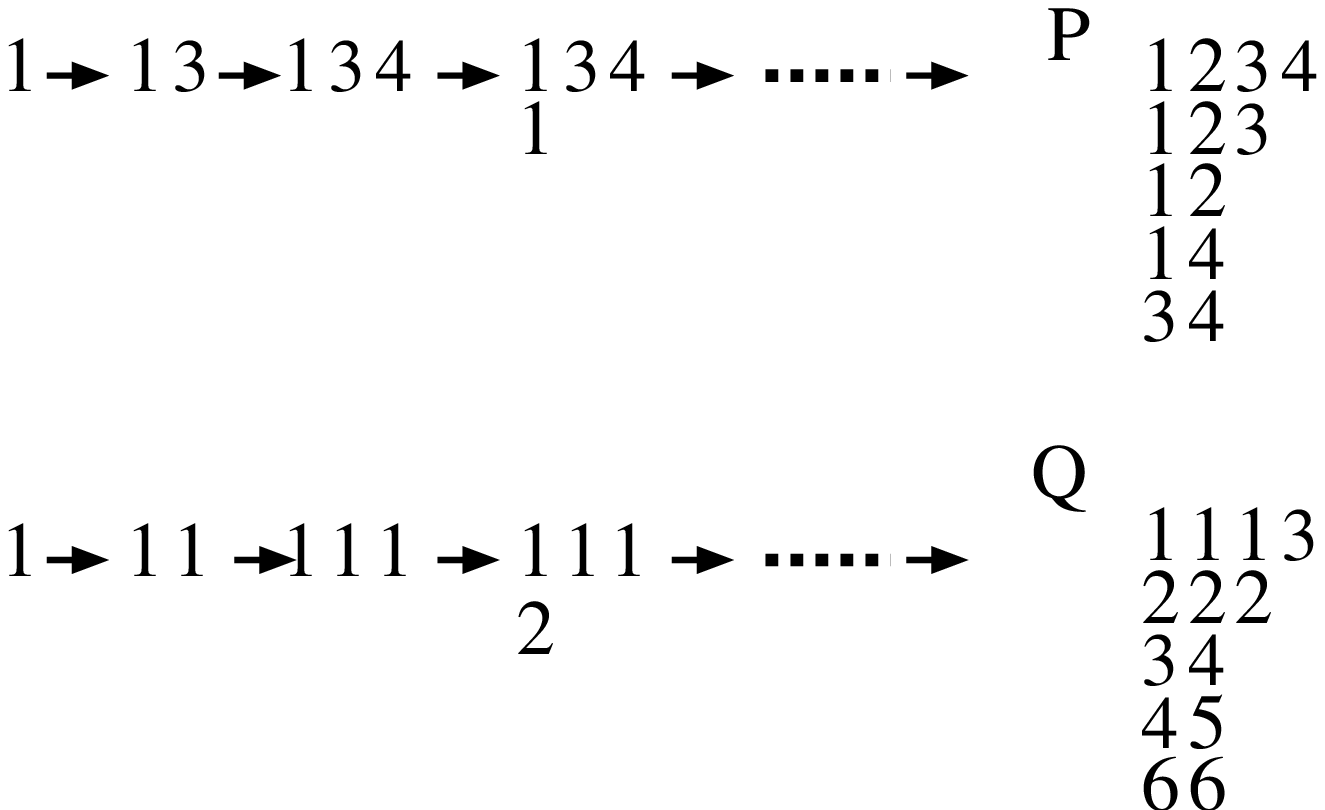}}}
\end{picture}

\vspace{10mm}

\noindent
Fig.~7

\begin{picture}(400,171)
\put(10,72){$\lambda(1,4)$}
\put(65,72){$\lambda(2,4)$}
\put(125,72){$\lambda(3,4)$}
\put(200,72){$\lambda(4,4)$}
\put(280,72){$\lambda(5,4)$}
\put(362,72){$\lambda(6,4)$}
\put(10,0){\scalebox{0.6}{\includegraphics{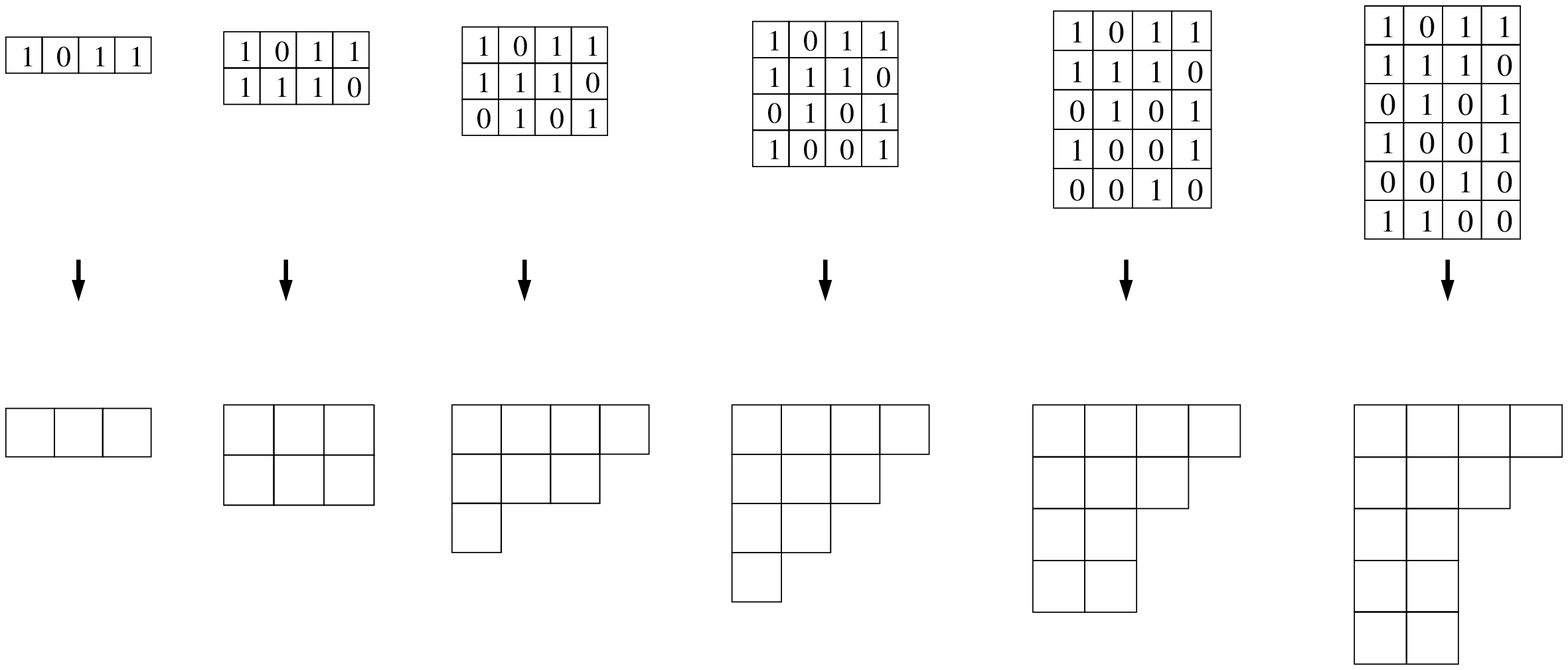}}}
\end{picture}

\noindent
Fig.~8

\begin{picture}(400,171)
\put(0,160){(a)}
\put(220,160){(b)}
\put(110,0){$s$}
\put(330,0){$s$}
\put(0,5){\scalebox{0.6}{\includegraphics{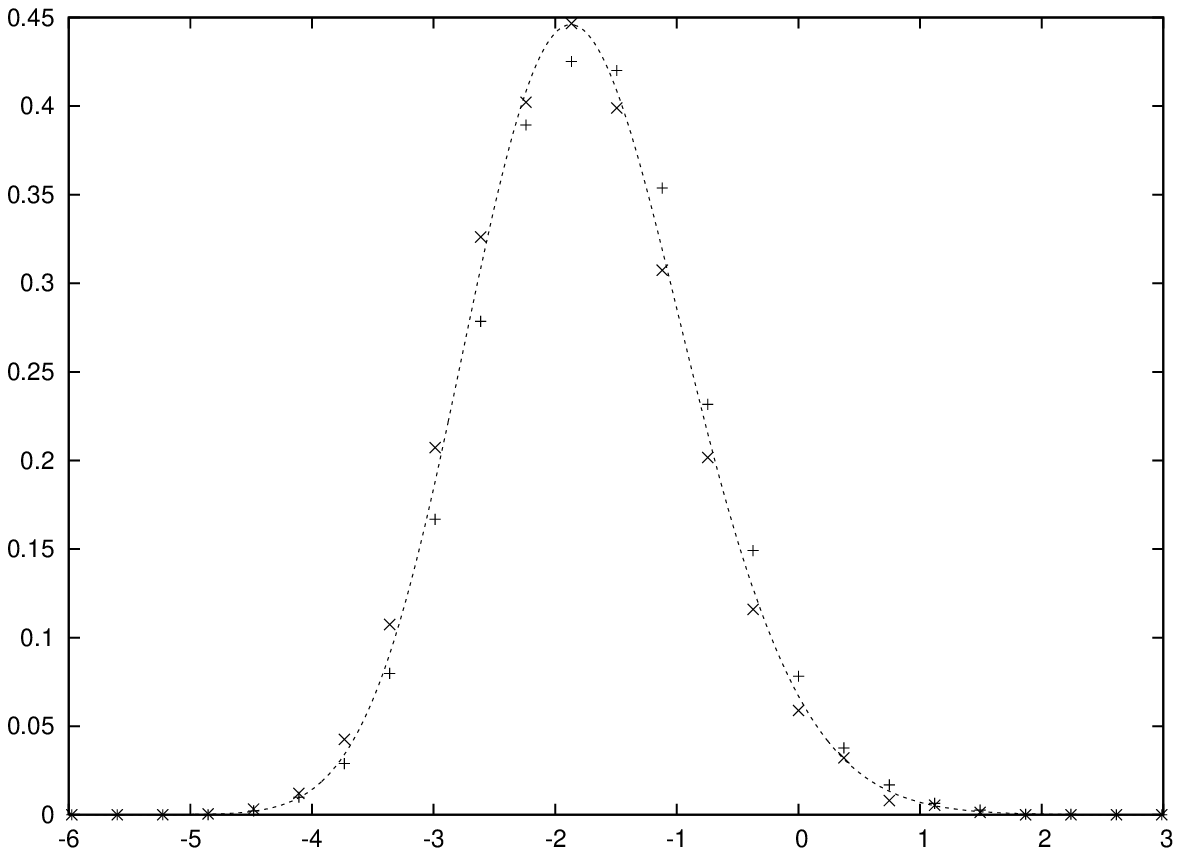}}}
\put(220,5){\scalebox{0.6}{\includegraphics{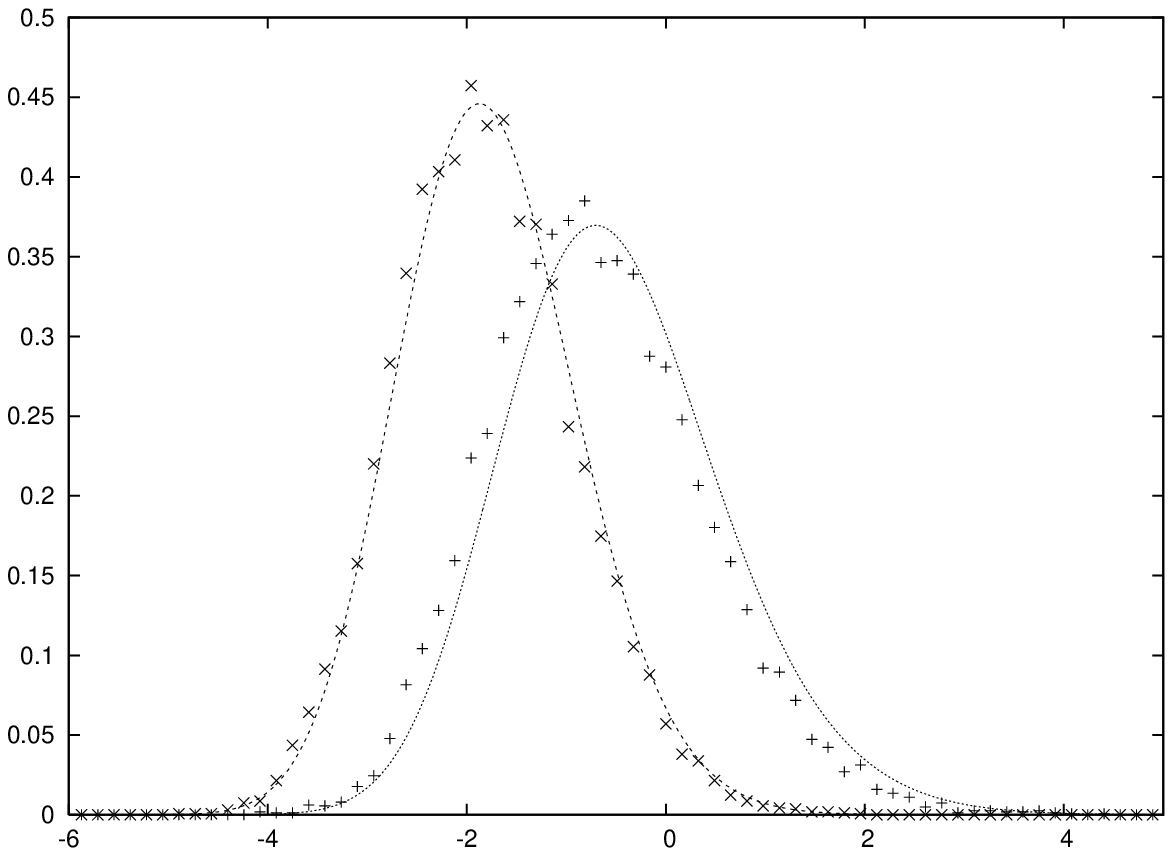}}}
\end{picture}

\begin{picture}(400,171)
\put(0,160){(c)}
\put(110,0){$s$}
\put(0,5){\scalebox{0.6}{\includegraphics{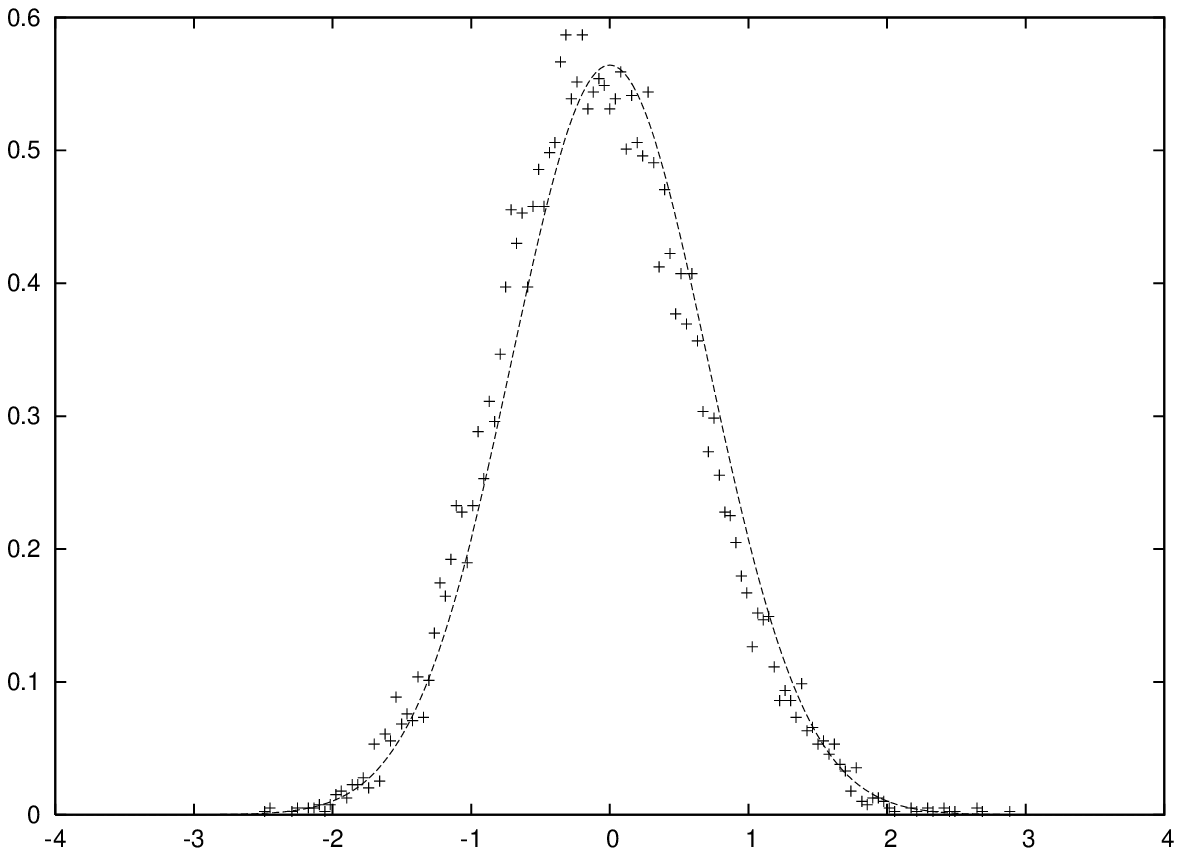}}}
\end{picture}

\begin{thebibliography}{10}
\bibitem{Li1985}
T.~M.~Liggett,
\newblock {\em Interacting Particle Systems},
\newblock Splinger-Verlag, New York, 1985.

\bibitem{Li1999}
T.~M.~Liggett,
\newblock {\em Stochastic Interacting Systems: Contact, Voter, and 
Exclusion Processes},
\newblock Splinger-Verlag, New York, 1999.

\bibitem{Sp1991}
H.~Spohn,
\newblock {\em Large Scale Dynamics of Interacting Particles},
\newblock Splinger-Verlag, New York, 1991.

\bibitem{ScZi1994}
B.~Schmittmann and R.~K.~P.~Zia,
\newblock {\em Statistical mechanics of driven diffusive systems},
\newblock{in C.~Domb and J.~Lebowitz eds., {\em Phase Transitions and 
Critical Phenomena,}} 17,
\newblock Academic, London, 1994.

\bibitem{Sc2001}
G.~M.~Sch{\"u}tz,
\newblock {\em Solvable Models for Many-Body Systems Far from 
Equilibrium} 
\newblock{in C.~Domb and J.~Lebowitz eds., {\em
Phase Transitions and Critical Phenomena}}, 19: 1--251, 
\newblock Academic, London, 2001.

\bibitem{DEHP1993}
{B.~Derrida, M.~R.~Evans, V.~Hakim, and V.~Pasquir},
\newblock{Exact solution of a 1D exclusion model using a matrix 
formulation},
\newblock{\em J. Phys. A.}, 26: 1493--1517, 1993.

\bibitem{Sa1999}
{T.~Sasamoto},
\newblock{One-dimensional partially asymmetric simple exclusion process 
with open boundaries: orthogonal polynomials approach},
\newblock{\em J. Phys. A.}, 32: 7109--7131, 1999.

\bibitem{USW2004}
{M.~Uchiyama, T.~Sasamoto, and M.~Wadati},
\newblock{Asymmetric simple exclusion process with open boundaries and 
Askey-Wilson polynomials},
\newblock{\em J. Phys. A.}, 37: 4985--5002, 2004.

\bibitem{Ki1986}
C.~Kipnis,
\newblock{Central limit theorem for infinite series of queues and 
applications to simple exclusion},
\newblock{\em Ann. Prob.}, 14: 397--408, 1986.

\bibitem{vB1991}
H.~van Beijeren,
\newblock{Fluctuation in the motions of mass and of patterns in 
one-dimensional hopping model},
\newblock{\em J. Stat. Phys.}, 63: 47--58, 1991.

\bibitem{MB1991}
S.~N.~Majumdar and M.~Barma,
\newblock{Tag diffusion in driven systems, growing surfaces, and 
anomalous fluctuations},
\newblock{\em Phys. Rev. B}, 44: 5306--5308, 1991.

\bibitem{KPZ1986}
M. Kardar, G. Parisi, and Y. C. Zhang,
\newblock{Dynamic scaling of growing interfaces},
\newblock{\em Phys. Rev. Lett.}, 56: 889--892, 1986.

\bibitem{GS1992}
{L.-H. Gwa and H. Spohn},
\newblock{Six-vertex model, roughened surfaces, and an asymmetric 
spin Hamiltonian},
\newblock{\em Phys. Rev. Lett.}, 68: 725--728, 1992.

\bibitem{Ki1995}
D.~Kim,
\newblock{Bethe ansatz solution for crossover scaling functions of the
asymmetric XXZ chain and the Kardar-Parisi-Zhang-type growth model},
\newblock{\em Phys. Rev. E}, 52: 3512--3524, 1995.

\bibitem{Se1997}
T.~S{e}pp{\"a}l{\"a}inen,
\newblock{A scaling limit for queues in series},
\newblock{\em Ann. Appl. Prob.}, 7: 855--872, 1997.

\bibitem{BDJ1999}
{J. Baik, P. A. Deift, and K. Johansson},
\newblock{On the distribution of the length of the longest increasing
subsequence in a random permutation},
\newblock{\em J. Amer. Math. Soc.}, 12: 1119--1178, 1999.

\bibitem{Jo2000}
K.~Johansson,
\newblock{Shape fluctuations and random matrices,}
\newblock{\em Commun. Math. Phys.}, 209: 437--476, 2000.

\bibitem{BR2001a}
J.~Baik and E.~M. Rains,
\newblock{Algebraic aspects of increasing subsequences},
\newblock{\em Duke Math. J.}, 109: 1--65, 2001.

\bibitem{BR2001b}
J.~Baik and E.~M. Rains,
\newblock{The asymptotics of monotone subsequences of involutions},
\newblock{\em Duke Math. J.}, 109: 205--281, 2001.

\bibitem{BR2001c}
J.~Baik and E.~M. Rains,
\newblock{Symmetrized random permutations},
\newblock in P.~M. Bleher and A.~R. Its, eds., 
{\em Random Matrix Models and Their Applications}, 1--29, 
Cambridge University Press, 2001.

\bibitem{TW1994}
C.~A. Tracy and H.~Widom,
\newblock{Level-spacing distributions and the Airy kernel},
\newblock{\em Commun. Math. Phys.}, 159: 151--174, 1994.

\bibitem{NS2004}
T.~Nagao and T.~Sasamoto,
\newblock{Asymmetric simple exclusion process and modified random
matrix ensembles},
\newblock{\em Nucl. Phys. B}, 699: 487--502, 2004.

\bibitem{RS2005}
A.~R{\'a}kos and G.~M.~Sch{\"u}tz,
\newblock{Current distribution and random matrix ensembles 
for an integrable asymmetric fragmentation process},
\newblock{\em J. Stat. Phys.}, 118: 511--530, 2005.

\bibitem{TW1996}
C.~A. Tracy and H.~Widom,
\newblock{On orthogonal and symplectic matrix ensembles},
\newblock{\em Commun. Math. Phys.}, 177: 727--754, 1996.

\bibitem{PS2002a}
M.~Pr{\"a}hofer and H.~Spohn,
\newblock{Current fluctuations for the totally asymmetric simple 
exclusion process},
\newblock In V.~Sidoravicius, ed., {\em In and out of equilibrium, 
{\rm Series:} Progress in Probability}, 51:  185--204, Birkh{\"a}user, 
2002.

\bibitem{Sa2005}
T.~Sasamoto,
\newblock{Spatial correlations of the 1D KPZ surface on a
flat substrate},
\newblock{\em J. Phys. A}, 38: L549--L556, 2005.

\bibitem{S1997}
G.~M.~Sch{\"u}tz,
\newblock{Exact solution of the master equation for the 
asymmetric exclusion process},
\newblock{\em J. Stat. Phys.}, 88: 427--445, 1997.

\bibitem{Se1998}
T.~S{e}pp{\"a}l{\"a}inen,
\newblock{Exact limiting shape for a simplified model of first-passage 
percolation on the plane},
\newblock{\em Ann. Prob.}, 26: 1232--1250, 1998.

\bibitem{J2001}
K.~Johansson,
\newblock{Discrete orthogonal polynomial and the Plancherel measure},
\newblock{\em Ann. Math.}, 153: 259--296, 2001.

\bibitem{GTW2001}
J. Gravner, C. A. Tracy, and H. Widom,
\newblock{Limit theorems for height fluctuations in a class of discrete space
and time growth models},
\newblock {\em J. Stat. Phys.}, 102: 1085--1132, 2001.

\bibitem{GTW2002a}
J. Gravner, C. A. Tracy, and H. Widom,
\newblock{A growth model in a random environment},
\newblock {\em Ann. Prob.}, 30: 1340--1369, 2002.

\bibitem{OC2003a}
N.~O'Connell,
\newblock{Conditioned random walk and the RSK correspondence},
\newblock{\em J. Phys. A}, 36: 3049--3066, 2003. 

\bibitem{OC2003b}
N.~O'Connell,
\newblock{A path-transformation for random walks and the 
Robinson-Schensted correspondence},
\newblock{\em Trans. Am. Math. Soc.}, 355: 3669--3097, 2003.

\bibitem{BO2006a}
A.~Borodin and G.~Olshanski,
\newblock{Stochastic dynamics related to Plancherel measures
on partitions},
\newblock{In V.~A.~Kaimanovich and A.~Lodkin, eds, 
\em{Representation Theory, Dynamical Systems, and Asymptotic 
Combinatorics (American Mathematical Society Translations Series 2,
Vol. 217)}}, 9--22, 2006. 

\bibitem{BO2006b}
A.~Borodin and G.~Olshanski,
\newblock{Markov processes on partitions},
\newblock{\em Probab. Theory. Relat. Fields}, 135: 84--152, 2006. 

\bibitem{OkRe2003}
A.~Okounkov and N.~Reshetikhin,
\newblock{Correlation function of Schur process with application 
to local geometry of a random 3-dimensional Young diagram,}
\newblock{\em J. Amer. Math. Soc.}, 16: 581--603, 2003.

\bibitem{BoRa2006}
A.~Borodin and E.~M.~Rains,
\newblock{Eynard-Mehta theorem, Schur process, and their 
Pfaffian analogs,}
\newblock{\em J. Stat. Phys.}, 121: 291--317, 2006.

\bibitem{Jo2003}
K.~Johansson,
\newblock{Discrete polynuclear growth and determinantal processes},
\newblock{\em Commun. Math. Phys.}, 242: 277--329, 2003.

\bibitem{BFP2006p}
A.~Borodin, P.~L.~Ferrari, M.~Pr{\"a}hofer,
\newblock{Fluctuations in the discrete TASEP with periodic initial 
configurations and the Airy$_1$ process},
\newblock math-ph/0611071.

\bibitem{PP2006}
A.~M.~Povolotsky and V.~B.~Priezzhev,
\newblock{Determinant solution for the totally asymmetric exclusion
process with parallel update,}
\newblock{\em J. Stat. Mech.}, P07002, 2006.

\bibitem{BFPS2006p}
A.~Borodin, P.~L.~Ferrari, M.~Pr{\"a}hofer, and T.~Sasamoto,
\newblock{Fluctuation properties of the TASEP with periodic 
initial configuration},
\newblock math-ph/0608056.

\bibitem{BBAP2005}
{J.~Baik, G.~Ben Arous, and S.~P{\'e}ch{\'e}},
\newblock{Phase transition of the largest eigenvalue for 
non-null complex sample covariance matrices},
\newblock{\em Ann. Prob.}, 33: 1643--1697, 2005.

\bibitem{Ba2006}
{J.~Baik},
\newblock{Painlev{\'e} formulas of the limiting distributions for 
non-null complex sample covariance matrices},
\newblock{\em Duke Math. J.}, 133: 205--235, 2006.

\bibitem{RS2006}
A.~R{\'a}kos and G.~M.~Sch{\"u}tz,
\newblock{Bethe ansatz and current distribution for the TASEP with 
particle-dependent hopping rates},
\newblock{\em Mar. Pro. Relat. Fields}, 12: 323--334, 2006.

\bibitem{E1996}
M.~R.~Evans,
\newblock{Bose-Einstein condensation in disordered exclusion models
and relation to traffic flow,}
\newblock{\em Europhys. Lett.}, 36: 13--18, 1996.

\bibitem{E1997}
M.~R.~Evans,
\newblock{Exact steady states of disordered hopping particle models
with parallel and ordered sequential dynamics,}
\newblock{\em J. Phys. A.}, 30: 5669--5685, 1997. 

\bibitem{SeKr1999}
T.~S{e}pp{\"a}l{\"a}inen and J.~Krug,
\newblock{Hydrodynamics and platoon formation for a totally 
asymmetric exclusion model with particlewise disorder},
\newblock{\em J. Stat. Phys.}, 95: 525--567, 1999.

\bibitem{AAR1999}
G.~E.~Andrews, and R.~Askey, and R.~Roy,
\newblock{\em Special Functions},
\newblock{Cambridge University Press}, 1999.

\bibitem{BO2006c}
A.~Borodin, and G.~Olshanski,
\newblock{Asymmetrics of Plancherel-type random partitions},
\newblock{\em J. Albebra}, 2007, doi:10.1016/j-jalgebra.2006.10.039,
math.PR/0610240.

\bibitem{FNH1999}
P.~J.~Forrester, T.~Nagao, and G.~Honner,
\newblock{Correlations for the orthogonal-unitary and symplectic transitions at
  the hard and soft edges},
\newblock{\em Nucl. Phys. B}, 553: 601--643, 1999.

\bibitem{Ma1994}
A.~M.~S. Mac\^edo,
\newblock{Universal parametric correlations at the soft edge of 
spectrum of random matrix ensembles},
\newblock{\em Europhys. Lett.}, 26: 641--646, 1994.

\bibitem{PS2002b}
M.~Pr{\"a}hofer and H.~Spohn,
\newblock{Scale invariance of the {PNG} droplet and the Airy process,}
\newblock{\em J. Stat. Phys.}, 108: 1071--1106, 2002.

\bibitem{Dy1962}
F.~J. Dyson,
\newblock{A Brownian-motion model for the eigenvalues of a random 
matrix},
\newblock{\em J. Math. Phys.}, 3: 1191--1198, 1962.

\bibitem{IS2004}
T. Imamura and T. Sasamoto,
\newblock{Fluctuations of the one-dimensional polynuclear growth model 
with external sources},
\newblock{\em Nucl. Phys. B}, 699: 503--544, 2004.

\bibitem{IS2005}
T.~Imamura and T.~Sasamoto,
\newblock{Polynuclear growth model with external source and 
random matrix model with deterministic source},
\newblock{\em Phys. Rev. E}, 71: 041696, 2005.

\bibitem{BR2000}
J.~Baik and E.~M. Rains,
\newblock{Limiting distributions for a polynuclear growth model with 
external sources},
\newblock{\em J. Stat. Phys.}, 100: 523--541, 2000.

\bibitem{Fo2000p}
P.~J.~Forrester,
\newblock{Painlev\'e transcendent evaluation of the scaled distribution
of the smallest eigenvalue in the Laguerre orthogonal and symplectic 
ensembles},
nlin.SI/0005064.

\bibitem{DF2006p}
P.~Desrosiers and P.~J.~Forrester,
\newblock{Asymptotic correlations for Gaussian and Wishart 
matrices with external source},
\newblock{\em Int. Math. Res. Not.}, 2006: 27395, 2006. 

\bibitem{EM1999}
B.~Eynard and M.~L.~Mehta,
\newblock{Matrices coupled in a chain: I. Eigenvalue correlations},
\newblock{\em J. Phys. A}, 31: 4449--4456, 1998.

\bibitem{St1999}
R.~P.~Stanley,
\newblock{\em Enumerative Combinatorics Volume 2},
\newblock{Cambridge University Press}, 1999.

\bibitem{Ful1999}
W.~Fulton,
\newblock{\em Young tableaux},
\newblock{Cambridge University Press}, 1999.

\bibitem{Fe2003}
P.~L.~Ferrari and H.~Spohn,
\newblock{Step fluctuations for a faceted crystal,}
\newblock{\em J. Stat. Phys.}, 113: 1--46, 2003.

\bibitem{Jo2005}
K.~Johansson,
\newblock{The arctic circle boundary and the Airy process,}
\newblock{\em Ann. Prob.}, 33: 1--30, 2005.
\end{thebibliography}
\end{document}